\documentclass[superscriptaddress, preprint, amsmath, amssymb, aps, pra, floatfix, nofootinbib,11pt]{revtex4-1}
\usepackage[utf8x]{inputenc}
\usepackage{amsmath}

\usepackage{graphicx}
\usepackage{float}
\usepackage{wrapfig}

\usepackage{dcolumn}
\usepackage{bm}
\usepackage[colorlinks]{hyperref}
\usepackage{natbib}
\usepackage{subfigure}

\usepackage{xcolor}
\usepackage{soul}

\newcommand{\ie}{\emph{i.e.}, }
\newcommand{\eg}{\emph{e.g.}, }
\newcommand{\nn}{\nonumber}
\newcommand{\nnl}{\nonumber\\}

\newcommand{\ket}[1]{\left| #1 \right\rangle}
\newcommand{\bra}[1]{\left\langle #1 \right|}

\newcommand{\comm}[2]{\left[ #1, #2 \right]}

\newcommand{\expn}[1]{{\rm e}^{#1}}

\newcommand{\dg}{{^{\dagger}}}

\newcommand{\ax}{{\rm aux}}

\newcommand{\iop}{\hat{1}}
\newcommand{\sop}[2]{\left\lceil#1\big|#2\right\rfloor}
\newcommand{\sket}[2]{\Big[\ket{#1}\otimes \ket{#2}^*\Big]}
\newcommand{\sbra}[2]{\Big[\bra{#1}\otimes \bra{#2}^*\Big]}
\newcommand{\scomm}[1]{\left(\sop{#1}{\iop}-\sop{\iop}{#1}\right)}
\newcommand{\sacomm}[1]{\left(\sop{#1}{\iop}+\sop{\iop}{#1}\right)}

\begin{document}

\title{Quantum simulation of weak-field light-matter interactions}

\author{Steve M. Young}
\email{syoung1@sandia.gov}
\affiliation{Quantum Algorithms and Applications Collaboratory, Sandia National Laboratories, Albuquerque, New Mexico 87185, USA}
\author{Hartmut H\"affner}
\affiliation{Department of Physics, University of California, Berkeley, California 94720, USA}
\author{Mohan Sarovar}
\email{mnsarov@sandia.gov}
\affiliation{Quantum Algorithms and Applications Collaboratory, Sandia National Laboratories, Livermore, California 94550, USA}

\date{\today}

\begin{abstract}
Simulation of the interaction of light with matter, including at the few-photon level, is important for understanding the optical and optoelectronic properties of materials,
and for modeling next-generation non-linear spectroscopies that use entangled light.
At the few-photon level the quantum properties of the electromagnetic field must be accounted for with a quantized treatment of the field, and then such simulations quickly become intractable, especially if the matter subsystem must be modeled with a large number of degrees of freedom, as can be required to accurately capture many-body effects and quantum noise sources. Motivated by this we develop a quantum simulation framework for simulating such light-matter interactions on platforms with controllable bosonic degrees of freedom, such as vibrational modes in the trapped ion platform. The key innovation in our work is a scheme for simulating interactions with a continuum field using only a few discrete bosonic modes, which is enabled by a Green's function (response function) formalism. We develop the simulation approach, sketch how the simulation can be performed using trapped ions, and then illustrate the method with numerical examples. Our work expands the reach of quantum simulation to important light-matter interaction models and illustrates the advantages of extracting dynamical quantities such as response functions from quantum simulations.
\end{abstract}

\maketitle

\section{Introduction}
The fundamental physics of light-matter interactions was conceived more than half a century ago with the formulation of quantum electrodynamics \cite{Cohen-Tannoudji_1992, Cohen-Tannoudji_1997}, followed by the distillation of a theory of quantum optics \cite{Walls_Milburn_2007, Garrison_Chiao_2008}. Despite this, new surprising phenomena in the realm of how light interacts with matter are still being uncovered today. This is especially true in regimes of light-matter interaction where the electromagnetic field and the material system it interacts with must both be treated quantum mechanically -- \ie where semiclassical approximations break down. Examples of recent results in this area are the revelation that a single photon can be jointly absorbed by two atoms given the right conditions \cite{garziano_one_2016}, and the establishment of the fundamental limits and trade-offs present in building detectors for single or few photons \cite{young_fundamental_2018,young_general_2018, proppNonlinearAmplificationImproved2019,young_design_2020}.

There are also many applications that benefit from accurate modeling and simulation of the interaction of weak fields with complex material systems, including: (i) understanding and mimicking light absorption by photosynthetic organisms, where it is a challenge to understand the mechanisms by which the initial stages of photosynthesis can be efficient, even in the weak illumination conditions that many organisms live in \cite{Cook_Ko_Whaley_2021, Ko_Cook_Whaley_2021}, (ii) interpretation of recently developed nonlinear spectroscopies that use entangled states of few photons, which have the potential to provide unprecedented resolution of electronic, molecular, and condensed phase dynamics \cite{Schlawin_Mukamel_2013,Dorfman_Schlawin_Mukamel_2016}, and (iii) designing next-generation nanoscale engineered photodetectors that can tailor the interaction dynamics between light and matter \cite{young_fundamental_2018,young_general_2018, proppNonlinearAmplificationImproved2019,young_design_2020}.
In such settings, where exotic light fields like single photon wavepackets interact with nanoscale structured materials and molecules, one typically has a quantum many-body model description of the physics that cannot be solved exactly and is also intractable to solve numerically on a computer. Tractable semiclassical approximations that are suitable in other settings often produce inaccurate predictions in the settings described above. This model complexity is an obstacle to many important phenomena arising from weak-field light-matter interactions. Additionally, experiments themselves can be difficult to perform due to the need to prepare exotic quantum states of light.

These obstacles provide the motivation for the quantum simulation technique we present in this paper. Quantum simulators are nascent hardware platforms that have the potential to transform the landscape of what is tractable for simulation of physics and chemistry models. This technique can be applied to any problem involving the interaction of matter with weak field states of light and can be used to study the coherent dynamics of such systems. Additionally, this technique can be generalized to model other types of non-locally and/or time-dependently coupled baths such as phononic wavepackets.  We focus on analog quantum simulation platforms, where the underlying physics of the system to be simulated is encoded into the  Hamiltonian of a tunable system, who's dynamics then naturally carries out the simulation task.
Trapped ions are one of the leading platforms for analog quantum simulation due to the long coherence times, high degree of controllability, and scalability afforded by the platform
\cite{blatt_quantum_2012}.
In fact, recent demonstrations using trapped ions have included simulation of models of quantum magnetism
\cite{islam_emergence_2013},
of many-body localization phenomena \cite{smith_many-body_2016}, of discrete time crystals \cite{zhang_observation_2017}, and of energy transfer phenomena \cite{gorman_engineering_2018}.

\begin{figure}
	\centering
	\includegraphics[width=.8\columnwidth]{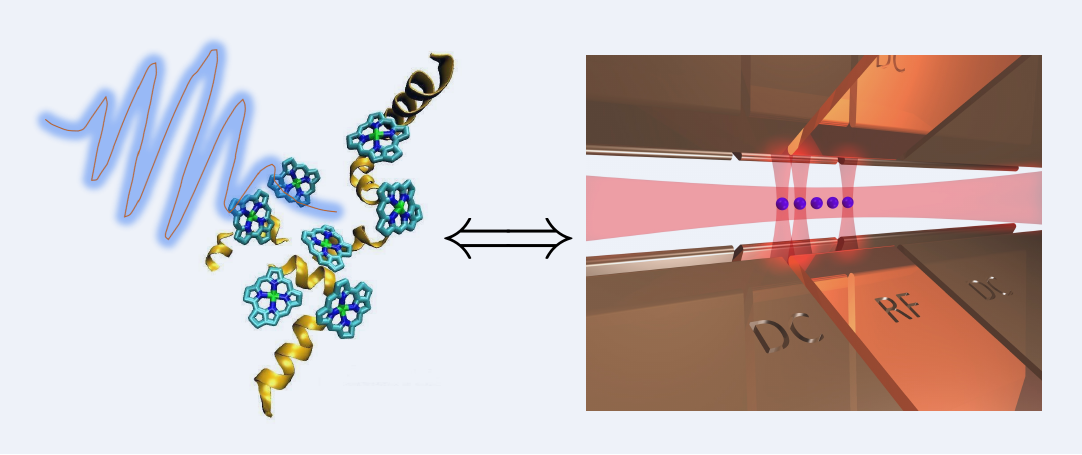}
	\caption{We develop a quantum simulation approach that enable platforms with controllable bosonic degrees of freedom, such as the trapped ion system shown on the right, to simulate the dynamics of matter, such as the chromophoric complex on the left, in response to illumination by extremely weak light fields. Such dynamics is relevant for understanding the behavior of photodetectors, photovoltaics, biochemical systems such as photosynthetic light harvesting complexes, and for modeling new weak-field spectroscopy experiments. \label{fig:fig}}
\end{figure}

In this work we study how light-matter interactions between exotic weak light fields and fundamental matter degrees of freedom can be simulated using ions trapped in linear radio-frequency (RF) traps, see Fig. \ref{fig:fig}. A key aspect of the quantum simulation model we propose is that the electromagnetic field is modeled by the quantized vibrational degrees of freedom in a trapped ion system. Due to the high degree of controllability of the trapped ion system, almost arbitrary states of the vibrational modes can be engineered, thus allowing simulation of interactions with exotic electromagnetic field modes that would be difficult to prepare, especially at optical frequencies. While electromagnetic field modes and vibrational modes are both bosonic degrees of freedom, an immediate obstacle to a simulation of the former with the latter is that while a general electromagnetic field is described by a continuum of harmonic modes, any simulation platform only contains a small, discrete number of vibrational modes. To overcome this difficulty, we formulate a novel form of simulation that proceeds via reconstruction of response functions from ensembles of quantum simulation experiments, and analyze the simulation cost in terms of the model being simulated. We also provide a detailed description and analysis of the implementation of our simulation protocol on a trapped-ion quantum simulator.

In the more general context, the approach we sketch is an example of using quantum simulation for calculation of physically relevant response functions \cite{Pedernales_2014, Xin_Pedernales_Lamata_Solano_Long_2017, Roggero_Carlson_2019, Kosugi_Matsushita_2020}. Although we present a formulation of the approach specific to analog trapped-ion quantum simulators, it is suitable for other hardware platforms with bosonic degrees of freedom (\eg circuit-QED \cite{Blais_Grimsmo_2021} or cavity-QED \cite{Haroche_Raimond_2006}), and there is a natural digitization of the approach, \eg through Trotterization, which we will comment on in Sec. \ref{sec:scaling}. Quantum simulation is widely believed to have potential to enable simulations that are intractable on conventional computers. Our methods introduce the simulation of weak-field light matter interactions into the quantum simulation toolbox, and as quantum simulation capabilities scale, our methods could be applied to model the interaction of weak electromagnetic fields with materials and molecules that are too complex for conventional simulation and modeling tools.

We note that there have been previous proposals to simulate vibrational degrees of freedom in molecules with bosonic degrees of freedom in trapped-ion systems \cite{MacDonell_2021} and to utilize quantum optical networks to emulate and sample vibronic spectra of molecules \cite{Shen_2018, Wang_2020}. To our knowledge, our proposal is the first to show that simulation of quantized states of electromagnetic fields is also possible with trapped-ion bosonic modes.

The remainder of the paper is structured as follows. In Sec. \ref{sec:ions} we present a brief overview of trapped-ion physics and the type of Hamiltonian models that can be engineered on this platform. Sec. \ref{sec:interaction} presents the light-matter interaction and dynamics that we wish to simulate, and then in Sec. \ref{sec:simulation} we detail our response function approach for simulating these dynamics on a platform like trapped ions with a fixed number of stationary bosonic modes. Sec. \ref{sec:eg} presents examples illustrating the scheme, with numerical simulations. Then in Sec. \ref{sec:scaling} we discuss important considerations when scaling the proposed simulation scheme to large systems, and in Sec. \ref{sec:other} we discuss application of our response function approach to quantum simulation to models beyond light-matter interactions. Finally, Sec. \ref{sec:disc} concludes with a summary of contributions made in this work.

\section{Trapped ion Hamiltonians}
\label{sec:ions}
The trapped-ion platform allows for realization of a rich set of quantum models with localized degrees of freedom, encoded in internal states of the ions, and distributed bosonic degrees of freedom, encoded in quantized motional modes of ion motion. In the following, we briefly review the building blocks of quantum simulation using trapped ions, with a particular focus on the physics and achievable Hamiltonian models that are relevant to our setting. We focus on linear ions chains of Ca$^+$ ions for concreteness, although  neither is an important restriction. We emphasize that this is a narrow review and refer the reader to Refs.
\cite{blatt_ion_2004, blatt_quantum_2012, bruzewicz_trapped-ion_2019}
for more comprehensive treatments.

We encode the ground state and optically connected excited state of point-like absorbers (\eg atoms, molecules) in two Zeeman sublevels of a Ca$^+$ ion. In particular, a commonly used sublevel encoding is $\ket{g} \rightarrow \ket{S_{1/2}(m_J = \frac{1}{2})}$ and $\ket{e} \rightarrow \ket{D_{5/2}(m_J=\frac{1}{2})}$, that uses states in the stable $S_{1/2}$ orbital and the metastable $D_{5/2}$ orbital \cite{gorman_engineering_2018}. These states will form the eigenbasis for localized degrees of freedom and in the following, when we write Pauli operators acting on localized degrees of freedom, they will be with respect to this basis. Note that it is possible to consider point-like absorbers with more than one excited state by encoding into other states in Zeeman sublevels of the ion, but we will restrict attention to single excited states here.

An ion chain with $n$ ions has $3n$ quantized motional modes (corresponding to collective motions of the $n$ ions under the influence of the overall trap potential and their mutual Coulomb repulsion) that can serve as modes of the electromagnetic field. This is less than the continuum required to accurately model electromagnetic (EM) fields but in the following we will construct a scheme for overcoming this obstacle.  All of the motional modes can in principle be coupled to the internal degrees of freedom of the ions through laser-induced interactions. The basic interaction that enables most of trapped ion quantum simulation is given by the following Hamiltonian that describes a single ion interacting with a laser field nearly resonant with the energy difference of the $\ket{e}$ and $\ket{g}$ states of the ion ($\hbar=1$ here and in the rest of the paper)
\cite{haeffner_quantum_2008}:
\begin{flalign}
	H^{\rm SI}(t) = \Omega \hat{\sigma}^+ e^{-i(\Delta t - \varphi)} \exp\left(i\eta \left[ \hat{a}e^{-i\omega_{\rm T} t} + \hat{a}\dg e^{i\omega_{\rm T}  t} \right]\right) + h.c., \nn
\end{flalign}
where $\Delta$ is the detuning of the laser from the atomic transition, $\varphi$ is the phase of the laser with respect to the atomic polarization, $\Omega$ is the amplitude of the laser field (expressed in terms of the Rabi frequency), $\omega_{\rm T} $ is the trap frequency. $\hat{\sigma}^+$ is the raising operator for the internal degrees of freedom, and $\hat{a}$ is the annihilation operator for the quantized motion of the ion around its equilibrium position along, for example, the $z$ axis. $\eta \equiv k_z \sqrt{\frac{1}{2m\omega_{\rm T} }}$ is the Lamb-Dicke parameter, with $k_z$ being the projection of the laser's wavevector along the $z$ direction. In the Lamb-Dicke limit, which describes the regime of small light-induced changes in momenta, or $\eta \sqrt{\langle (\hat{a}+\hat{a}\dg)^2\rangle} \ll 1$, this Hamiltonian can be approximated as
\cite{haeffner_quantum_2008}
\begin{flalign}
	H^{\rm SI}(t) &\approx \Omega \left( \hat{\sigma}^+ e^{-i(\Delta t - \varphi)} + \hat{\sigma}^- e^{i(\Delta t-\varphi)}\right) \nn \\
& + i\eta\Omega\left(\hat{\sigma}^+ e^{i(\Delta t - \varphi)} - \hat{\sigma}^- e^{i(\Delta t - \varphi)}\right)\left( \hat{a}e^{-i\omega_{\rm T}  t} + \hat{a}\dg e^{i\omega_{\rm T}  t}\right).
\label{eq:ion-mode}
\end{flalign}
Then, through choices of $\Delta$ and $\varphi$ one can engineer a wide variety of ion-mode interactions. The above description of the dynamics of a single ion generalizes easily to multiple ions, with the motion now being interpreted as modes of the collective motion of the ion chain.

In addition to the ion-mode interactions, we can also mediate interactions between the internal states of ions through their common interaction with motional modes. There are a few schemes for engineering such interactions, and the commonly used M{\o}lmer-S{\o}renson scheme -- which illuminates two ions with a bichromatic laser with frequencies $\omega_0 \pm (\nu + \delta)$, where $\omega_0$ is the energy difference between $\ket{g}$ and $\ket{e}$, $\nu$ is the frequency of a vibrational mode (usually an axial mode), and $\delta$ is a detuning -- generates an effective interaction between two ions of the form $\hat{\sigma}^x_1 \hat{\sigma}^x_2$.

Putting these ingredients together we write the family of Hamiltonians that can be engineered with trapped ions -- which we denote with the superscript TI -- as
\begin{flalign}
	\hat{H}^{\rm TI} &= \hat{H}_a^{\rm TI} + \hat{H}_m^{\rm TI} + \hat{H}_I^{\rm TI}, \quad \textrm{with},  \label{eq:gen_ion_ham}\\
	\hat{H}_a^{\rm TI} &= \sum_{j=1}^n \omega_0^j \hat{\sigma}^z_j + \sum_{\langle i,j \rangle} J_{ij} \hat{\sigma}^x_i \hat{\sigma}^x_j, \label{eq:ha} \\
	\hat{H}_m^{\rm TI} &= \sum_{k=1}^m \nu_k \hat{a}_k\dg \hat{a}_k  \\
	\hat{H}_I^{\rm TI} &= \sum_\alpha \kappa_\alpha (\hat{\sigma}^+_{j_\alpha} \hat{a}_{k_\alpha} + \hat{\sigma}^-_{j_\alpha} \hat{a}\dg_{k_\alpha}) + \sum_\alpha \lambda_\alpha \hat{\sigma}^z_{j_\alpha}(\hat{a}_{k_\alpha} + \hat{a}\dg_{k_\alpha}).
\end{flalign}
Here, $\hat{H}_a^{\rm TI}$ is the Hamiltonian for the internal states of the $n$ ions, representing localized degrees of freedom. $\omega_0^j$ is the local transition energy and can be tuned by AC Stark shifting the magnetic sublevels of each ion with laser beams that locally address the ions. The second term represents coupling between some subset of ions and are implemented via the M{\o}lmer-S{\o}renson interaction as mentioned above. By tuning the M{\o}lmer-S{\o}renson interaction this coupling between ions can be tuned to the rotating wave version where only the terms $\hat{\sigma}_i^+  \hat{\sigma}_j^- + \hat{\sigma}_i^-  \hat{\sigma}_j^+$ contribute. The magnitude of the couplings, $J_{ij}$, can be tuned via the intensities and detunings of the laser beams implementing these interactions.
We note that engineering fully tunable $J_{ij}$ interaction terms through the M\o lmer-S\o renson (MS) interaction requires coupling to one motional mode per interaction term, which must be accounted for when accounting for quantum simulation resources.
$\hat{H}_m^{\rm TI}$ is the Hamiltonian representing the $m \leq 3n$ motional modes used in the simulation. The harmonic frequencies $\nu_j$ are set by the trapping potential and number of ions in the trap but the effective mode frequencies that the ions interact with can often be tuned via how the interactions are engineered \cite{gorman_engineering_2018}. $\hat{H}_I^{\rm TI}$ represents the interactions between the ions and motional modes, and accounts for two types of interactions. The first sum represents a coherent exchange of energy and is implemented via the interaction discussed in Eq. (\ref{eq:ion-mode}) (by setting $\Delta=-\omega_{\rm T}$, or a red sideband drive). The second represents a shift in the energy of the ions that depends on the position of the motional mode and can be implemented via a bichromatic local addressing of the ion involved \cite{gorman_engineering_2018}. The index $\alpha$ in these sums represent the number of interactions, and $j_\alpha$, $k_\alpha$ index the ions and modes involved in an interaction. The interactions parameters, $\kappa_\alpha$ and $\lambda_\alpha$, can be tuned by adjusting the intensity of the local addressing laser beams. Importantly, one ion cannot participate in both types of interactions. We note that other types of Hamiltonian terms are also possible but the family of Hamiltonians described above will be sufficient for our purposes.

Having described the types of Hamiltonians that can be engineered on the trapped-ion platform we will now summarize the other two requirements for a quantum simulation, \emph{state preparation} and \emph{measurement}. There are a variety of mechanisms in trapped ion platforms for extracting entropy and preparing desired states of the internal and motional degrees of freedom with high fidelity. The internal states of the ions can be prepared in the $\ket{g}$ state through optical pumping \cite{leibfried_quantum_2003}. The states of the motional modes can be prepared in thermal states, including very low temperature thermal states, through a variety of cooling mechanisms.  The most common such cooling mechanism, resolved sideband cooling, proceeds by coupling a mode to an ion using the same interaction as shown in Eq. (\ref{eq:ion-mode}) and cooling the mode by dumping excess energy into the ion, which emits into an optical mode usually controlled by coupling the long-lived excited state of the ion to a fast decaying excited state \cite{leibfried_quantum_2003}. Each mode that is used in the simulation can be cooled sequentially. If many modes need to be initialized cooling via electromagnetically induced  transparency might be an attractive alternative.  We note that other cooling mechanisms (\emph{e.g.,} Doppler cooling) must first be applied to the ion chain before the ion motion is sufficiently cold for resolved sideband cooling to be effective. Measurement of the ion internal degrees of freedom is accomplished via the electron shelving technique, which selectively excites one of the encoding states (usually $\ket{e}$) to a short-lived higher lying state whose emission is monitored.

In the following sections we will develop a scheme for simulating light-matter interactions using the quantum simulation building blocks discussed above.

\section{Interaction of matter with weak fields}
\label{sec:interaction}
Matter interacting with light must be considered as interacting with all modes allowed by the confining geometry, so that the Hamiltonian of an arbitrary system in the rotating-wave approximation can be written as
\begin{flalign}
\hat{H}&=\hat{H}_M+\hat{H}_F+\hat{H}_{M-F}\nnl
\hat{H}&=\hat{H}_M+\sum_j \omega_j \hat{a}_j\dg \hat{a}_j+\sum_j i(\hat{L}\hat{a}_j\dg-\hat{L}\dg \hat{a}_j)\nn
\end{flalign}
where $\hat{H}_M$ is an arbitrary Hamiltonian for the matter subsystem, $\hat{a}_j$ is the annihilation operator for allowed mode $j$, and $\hat{L}$ describes the action of the light-matter interaction on the matter system (\eg promoting a two-level system from the ground state to an excited state and vice versa, in which case $L=\hat{\sigma}^-$).  In the event that the geometry of the space containing the matter is unconfined; \ie the matter exists in free-space in at least one direction, then the allowed modes form a continuum.  For the 1D case (for simplicity and clarity)  modes in this continuum can be represented by their frequency $\omega$ and we can write
\begin{flalign}
\hat{H}&=\hat{H}_M+\int d\omega\omega\hat{a}(\omega)\dg \hat{a}(\omega)+\int d\omega i(\hat{L}\hat{a}(\omega)\dg-\hat{L}\dg \hat{a}(\omega))\label{eq:Hc}
\end{flalign}
Solution of this Hamiltonian using a Markov approximation results in the emergence of the phenomenon of spontaneous emission, as first identified by Wigner and Weisskopf, as a direct consequence of the availability of a continuum of emission channels.

Consequently, this Hamiltonian cannot be straightforwardly described by one, or even a few, bosonic modes, as in the case of ion trap system of Sec. \ref{sec:ions}.
Nonetheless, as we will now show, it is possible to use the ion trap system to simulate light-matter interactions of the former kind.

\section{Simulation through response functions}
\label{sec:simulation}
First, we will rewrite Eq.~\eqref{eq:Hc} according to the quantum noise formalism.  This can be done by constructing new field operators.  With the field only in the interaction picture,
\begin{flalign*}
\hat{dB}_t&\approx dt\int d\omega \expn{-i(\omega-\omega_0) t}\hat{a}(\omega)\\
\hat{H}_{M-F}(t)dt&=i( \expn{i\omega_0 t}\hat{L}\hat{dB}_t\dg - \expn{-i\omega_0 t}\hat{L}\dg\hat{dB}_t),
\end{flalign*}
where $\omega_0$ represents a frequency that will be associated with the incoming pulse.
The new operators behave like noise operators, and obey the same statistics, such that \cite{Combes_Kerckhoff_Sarovar_2017}
\begin{flalign}
[\hat{dB}_t,\hat{dB}_{t'}\dg]&=\delta_{t,t'}dt\nnl
\hat{dB}_t\hat{dB}\dg_t &=\mathcal{O}(dt)\label{eq:dbt_rule}
\end{flalign}
 A monochromatic single photon pulse with frequency $\omega_0$ traversing the matter system with a broad temporal profile $\varepsilon(t)$ is constructed as
\begin{flalign}
	\ket{1_\varepsilon}=\int \varepsilon(t)\hat{dB}_t\dg \ket{0}\label{eq:pulse},
\end{flalign}
where $\ket{0}$ is the field vacuum.

The form of the transformed Hamiltonian, $\hat{H}_{M-F}(t)$, makes it seem like the system is only coupled to a single field mode. However, the behavior of the $\hat{dB}_t$ operators will result in different dynamics. In particular, due to Eq.~(\ref{eq:dbt_rule}) the interaction Hamiltonian provides an additional, nonunitary term at linear order that must be included when propagating the dynamics; to see this explicitly, note that the time evolution operator for a short interval is
\begin{flalign*}
	\hat{\Upsilon}(dt)&=\expn{-i\hat{H}dt}=1-i\hat{H}dt-\hat{H}^2dt^2/2+ \dots\\
	&=1-i(\hat{H}_M+\hat{H}_{M-F})dt-\frac{1}{2}\left[\hat{H}_M^2dt^2+\hat{H}_M(\hat{H}_{M-F}dt)dt+(\hat{H}_{M-F}dt)\hat{H}_Mdt+(H_{M-F}dt)^2\right] + ...\\
	&=1-i\hat{H}_Mdt-i\hat{H}_{M-F}(t)dt- \hat{L}\dg\hat{L}/2dt+\mathcal{O}(dt\hat{dB}_t)
\end{flalign*}
Due to the nonunitary contribution we will write the evolution in terms of the density matrix
\begin{flalign}
	\hat{\rho}(t+dt)=&\hat{\Upsilon}(dt)\hat{\rho}(t)\hat{\Upsilon}\dg(dt)\nnl
	=&\hat{\rho}(t)-\left(i\comm{\hat{H}_M+\hat{H}_{M-F}}{\hat{\rho}(t)} +\frac{1}{2}\{\hat{L}\dg\hat{L},\hat{\rho}(t)\}\right)dt\nnl
	&+ \hat{L}\hat{dB}_t\dg\hat{\rho}(t)\hat{L}\dg\hat{dB}_t+\mathcal{O}(dt\hat{dB}_t)\label{eq:density_ev}
\end{flalign}

We note that incoherent processes (\eg due to couplings to environments other than the EM field) can be included as additional terms, \eg of the Lindblad form $\mathcal{D}[\hat{X}](\hat{\rho})=\hat{X}\hat{\rho}\hat{X}\dg-\frac{1}{2}\left\{\hat{X}\dg\hat{X},\hat{\rho}\right\}$, however we will omit these presently for clarity.
We will also ``vectorize'' the density matrix, performing the transformations  $\hat{\rho}_{ij}=\rho_{ij}\ket{i}\bra{j}\rightarrow\bar{\rho}_{(ij)}=\rho_{ij}\ket{i}\otimes\ket{j}^*$
and $\hat{X}\hat{\rho}\hat{Y}\dg\rightarrow\hat{X}\otimes\hat{Y}^T\bar{\rho} \equiv \sop{\hat{X}}{\hat{Y}}\bar{\rho}$. This allows us to express superoperators acting on $\hat{\rho}$, such as commutation and the Linbladian $\mathcal{D}$, as linear operators acting on $\bar{\rho}$.
We can now integrate Eq.~\eqref{eq:density_ev} directly. Let
\begin{flalign}
	\bar{G}_0(t)&=\expn{\left(-i\scomm{\hat{H}_M}-\frac{1}{2}\sacomm{\hat{L}\dg\hat{L}}\right)t}\nnl
	\bar{F}_{\hat{dB}}(t)&=\sop{\hat{L}\hat{dB}_t\dg}{\hat{L}\hat{dB}_t\dg}-i\scomm{\hat{H}_{M-F}(t)dt}
	\label{eq:greens}
\end{flalign}
Then
\begin{flalign*}
	\bar{\rho}(t)&=\bar{G}(t-t_0)\bar{\rho}(t_0)\\
	\bar{G}(t)&=\bar{G}_0(t)+\int^t_{t_0} \bar{G}_0(t-t')\bar{F}_{\hat{dB}}(t')\bar{G}_0(t')\\
	&+\int^t_{t_0} \bar{G}_0(t-t')\bar{F}_{\hat{dB}}(t')\int^{t'}_{t_0} \bar{G}_0(t'-t'')\bar{F}_{\hat{dB}}(t'')\bar{G}_0(t'')+...
\end{flalign*}
Here, the interaction between the system and field (represented by $\bar{F}_{\hat{dB}}(t)$) is treated perturbatively and each term in this expansion corresponds to an interaction of a certain order.
For a system with the matter subsystem initially in the ground state and the field in a single photon state, \ie  $\bar{\rho}(t_0)=\sket{0_M,1_\varepsilon}{0_M,1_\varepsilon}$, this series truncates and we get
\begin{flalign}
	\bar{\rho}(t)&=\bar{\rho}(t_0)- \int^t_{t_0}dt'\bar{\mathcal{L}}_+(t,t')\sket{0_M,0_\varepsilon}{0_M,1_\varepsilon}\nnl
	&+\int^{t}_{t_0}dt\int^{t'}_{t_0} dt''\bar{\mathcal{L}}_+(t,t')\bar{\mathcal{L}}\dg_+(t',t'')\sket{0_M,0_\varepsilon}{0_M,0_\varepsilon}\nnl
	&+\int^{t}_{t_0}dt'\int^{t'}_{t_0}dt''\int^{t''}_{t_0} dt'''\sop{\hat{L}}{\hat{L}}\bar{\mathcal{L}}_+(t',t'')\bar{\mathcal{L}}\dg_+(t'',t''')\sket{0_M,0_\varepsilon}{0_M,0_\varepsilon}+h.c.\label{eq:vector_ev}
\end{flalign}
with $\bar{\mathcal{L}}_+(t,t')=\bar{G}_0(t-t')\varepsilon(t')\expn{-i\omega_0t'}\sop{\hat{L}\dg}{\iop}$ for compactness in order to emphasize the structure (note that by construction $\sop{\hat{L}\dg}{\iop}\dg=\sop{\iop}{\hat{L}\dg}$).  The reason for the early truncation of this series is that once an absorbed photon is re-emitted into the continuum field it cannot interact with the matter system again; \ie $d\hat{B}_t\dg\ket{0_\varepsilon}\ne\ket{1_\varepsilon}$ \footnote{We use the $\varepsilon$ subscript on the vacuum state for uniformity.}. This is the essence of spontaneous emission, given by the third term of the sum.

We consider now the case where the system interacts with a single stationary bosonic-mode as in the trapped-ion context. The Hamiltonian, in an interaction picture with respect to the mode's free Hamiltonian, is:
\begin{align}
	\hat{H}^{\rm sm} &= \hat{H}_M + \gamma(t) H^{\rm sm}_{M-F}\\
	\hat{H}^{\rm sm} &= \hat{H}_M + i \gamma(t) (\expn{i\omega_0t}\hat{L} \hat{a}\dg-\expn{-i\omega_0t}\hat{L}\dg \hat{a}),
	\label{eq:h_sm}
\end{align}
where the superscript ``sm" on this quantity and subsequent quantities indicates that a single mode is modeling the EM field.
This is simply related to the general trapped ion Hamiltonian described in Eq. (\ref{eq:gen_ion_ham}); $\hat{H}_M$ is the matter Hamiltonian engineered through $\hat{H}_a^{\rm TI}$ in Eq. (\ref{eq:gen_ion_ham}), we have restricted to one mode, so $k_\alpha$ is the same for all $\alpha$, and $\hat{L}=\sum_\alpha \kappa_\alpha\hat{\sigma}_-^{j_\alpha}$ is the engineered interaction between the matter subsystem (modeled by the ions) and the mode. Finally, we have set $\lambda_\alpha=0$. In addition, for reasons that will become clear shortly, we will include a Markovian decoherence term to the dynamics of the form $\mathcal{D}[\hat{Y}]$, where $\hat{Y}$ is an operator on the internal states of the matter subsystem.

Given this setup the unitary contribution to the infinitesimal time evolution is governed by
\begin{flalign*}
	\hat{U}^{\rm sm}(dt)&=1-i\left(\hat{H}_M+i\gamma(t)\left(\hat{L}\hat{a}\dg\expn{-i\omega_0t}-\hat{L}\dg \hat{a}\expn{i\omega_0t}\right)\right)dt+\mathcal{O}(dt^2).
\end{flalign*}
Then, with the dissipative contribution included,
\begin{flalign}
	\bar{G}_0^{\rm sm}(t)&=\expn{\left(-i\scomm{\hat{H}_M}-\frac{1}{2}\sacomm{\hat{Y}\dg\hat{Y}}\right)t}\nnl
	\bar{F}_{\hat{a}}(t)&=-i\gamma(t)\scomm{ H^{\rm sm}_{M-F}}+\sop{\hat{Y}}{\hat{Y}}, \label{eq:greens_single}
\end{flalign}
and
\begin{flalign*}
	\bar{\rho}^{\rm sm}(t)&=\bar{G}^{\rm sm}(t-t_0)\bar{\rho}^{\rm sm}(t_0)\\
	\bar{G}^{\rm sm}(t)&=\bar{G}_0^{\rm sm}(t)+\int^t_0 dt'\bar{G}_0^{\rm sm}(t-t')\bar{F}_{\hat{a}}(t')\bar{G}_0^{\rm sm}(t')\\
	&+\int^t_0 dt'\bar{G}_0^{\rm sm}(t-t')\bar{F}_{\hat{a}}(t')\int^{t'}_0 dt''\bar{G}_0^{\rm sm}(t'-t'')\bar{F}_{\hat{a}}(t'')\bar{G}_0^{\rm sm}(t'')+...
\end{flalign*}
so that the dynamics can be expanded in the light-matter interaction for $\bar{\rho}(t_0)=\sket{0_M,1_F}{0_M,1_F}$ as
\begin{flalign}
	\bar{\rho}^{\rm sm}(t)&=\bar{\rho}^{\rm sm}(t_0)-\int_{t_0}^tdt'\bar{\mathcal{L}}^{\rm sm}_+(t,t')\sket{0_M,0_F}{0_M,1_F}\nnl
	&+\int_{t_0}^tdt'\int_{t_0}^{t'}dt''\bar{\mathcal{L}}^{\rm sm}_+(t,t')\bar{\mathcal{L}}^{\dagger \rm sm}_+(t',t'')\sket{0_M,0_F}{0_M,0_F}\nnl
	&+\int_{t_0}^tdt'\int_{t_0}^{t'}dt''\int_{t_0}^{t''}dt'''\sop{\hat{Y}}{\hat{Y}}\bar{\mathcal{L}}^{\rm sm}_+(t',t'')\bar{\mathcal{L}}^{\dagger \rm sm}_+(t'',t''') \sket{0_M,0_F}{0_M,0_F}\nnl
	&-\int_{t_0}^tdt'\int_{t_0}^{t'}dt''\bar{\mathcal{L}}^{\rm sm}_-(t,t')\bar{\mathcal{L}}^{\rm sm}_+(t',t'')\sket{0_M,1_F}{0_M,1_F}\nnl
	&+\int_{t_0}^tdt'\int_{t_0}^{t'}dt''\int_{t_0}^{t''}dt'''\bar{\mathcal{L}}^{\rm sm}_-(t,t')\bar{\mathcal{L}}^{ \rm sm}_+(t',t'')\bar{\mathcal{L}}^{\dagger \rm sm}_+(t'',t''')\sket{0_M,1_F}{0_M,0_F}+h.c.+\dots\label{eq:s_ev}
\end{flalign}
with $\bar{\mathcal{L}}$ expressions defined similarly to the continuum mode case (\eg $\bar{\mathcal{L}}^{ \rm sm}_-(t,t')=\bar{G}_0^{\rm sm}(t-t')\gamma(t')\expn{i\omega_0t'}\sop{\hat{L}}{\iop}$).
We can now compare Eq.\eqref{eq:vector_ev} and Eq.\eqref{eq:s_ev}, and find two essential differences. Firstly, the series in the continuum mode case, Eq. \eqref{eq:vector_ev}, cuts off after (spontaneous) emission, but continues with an infinite number of coherent absorption/emission cycles in the single mode case, Eq. \eqref{eq:s_ev}. Unlike in the continuum case, an excitation emitted into the stationary mode can be reabsorbed by the matter system. A second, related, difference is that in the continuum case, the light-matter interaction leads to a decoherent effect captured by the real terms in the argument of $\bar{G}^0$ and the second term in the definition of $\bar{F}_{\hat{dB}}(t)$, whereas in the single mode case decoherent effects are generated by the added term $\hat{Y}$.

Thus, in order to reproduce the continuum mode case with the single mode system, we must suppress the additional terms of Eq. \eqref{eq:s_ev} and set $\hat{Y}=\hat{L}$. Unfortunately, neither of these tasks is straightforward to carry out.  For the first, the absorption/reemission dynamics of the single mode case are inherent in the system. For the second, since $\hat{Y}$ is engineered via an off-resonant drive between the relevant internal states of the ions and higher lying state(s) with short lifetimes, or through other dissipative mechanisms \cite{Barreiro_Blatt_2011}, it must be a local operator, unlike $\hat{L}$, which couples multiple ions to the field.  We shall deal with both of these issues in turn, and show how simulation of relevant continuum dynamics can still be achieved with the single mode system.

\subsection{Moving to response functions}
A way to overcome the difficulty of not having a continuum of modes and the resulting phenomenon of spontaneous emission is to take advantage of the time dependence of the coupling, $\gamma(t)$, in Eq. (\ref{eq:greens_single}) and move to a response function framework.

First we note that, the population of an excited matter state $i_M$ at time $t$, which is often the quantity of interest, is given by the second term of Eq. \eqref{eq:vector_ev} after tracing over the field state:
\begin{flalign}
	P^M(t) &\equiv \bra{i_M} {\rm Tr}_{F}\left[ \hat{\rho}(t) \right]\ket{i_M} \nnl
	&=\int^{t}_{t_0}dt'\int^{t'}_{t_0} dt''\varepsilon(t')\varepsilon(t'')\mathcal{G}_{i_M}(t-t',t'-t'')
	\label{eq:pop_conv}
\end{flalign}
where
\begin{flalign}
	\mathcal{G}_{i_M}(t_1,t_2)\equiv\sbra{i_M}{i_M}\expn{-i\omega_0t_2}\bar{G}_0(t_1)\sop{\hat{L}\dg}{\iop}\bar{G}_0(t_2)\sop{\iop}{\hat{L}\dg}\sket{0_M}{0_M} + c.c., \label{eq:2tGf}
\end{flalign}
so that the dynamical properties are captured by this two-time Green's function. We write such Green's functions as $\mathcal{G}_{i_M}(t_1,t_2)$, where the subscript $i$ denotes the state whose population is being measured.   If this quantity is known, then the dynamics in response to any wavepacket profile shape $\varepsilon(t)$ can be determined.  The problem can then be reduced to finding the response to delta-like inputs.

To simulate the delta function response with just a single stationary mode, we can consider modulating the field coupling to be nonzero only briefly and at given times. Consider a square ``pulse'' of width $t_\gamma$ and area $n_\gamma$, so that $\gamma(t)=\frac{n_\gamma}{t_\gamma}\left[\theta(t_1-t_\gamma/2)-\theta(t_1+t_\gamma/2)\right]$. We note that these pulses do not need to be strictly square in shape; we assume as such for illustrative purposes. When $\bar{G}_0(t)$ can be approximated as constant over the interval $t_\gamma$, we obtain for a pair of pulses at $t'$ and $t''$
\begin{flalign}
	&P_{\rm sm}^M(t-t',t'-t'')\equiv \bra{i_M} {\rm Tr}_{F} \left[\hat{\rho}^{\rm sm}(t)\right] \ket{i_M} \nnl
	&=n_\gamma^2\expn{i\omega_0(t''-t')}\sbra{i_M}{i_M}\bar{G}_0^{\rm sm}(t-t')\sop{\hat{L}\dg}{\iop}\bar{G}_0^{\rm sm}(t'-t'')\sop{\iop}{\hat{L}\dg} \sket{0_M}{0_M} +h.c.\nnl
	&~~+n_\gamma^2\sbra{i_M}{i_M}\left(\bar{G}_0^{\rm sm}(t-t')+\bar{G}_0^{\rm sm}(t-t'')\right)\sop{\hat{L}\dg}{\hat{L}\dg} \sket{0_M}{0_M}+\mathcal{O}(\left[\vert\vert \hat{L} \vert\vert^2 n_\gamma^2\right]^4).\nn
\end{flalign}
We can see that when $\vert\vert \hat{L} \vert\vert^2n^2_\gamma \ll 1$ the terms to linear order dominate. Setting $\hat{Y}=\hat{L}$ for now, and comparing to Eq.~(\ref{eq:2tGf}), letting $t_m\equiv t-t'$ and $t_{\rm int}\equiv t'-t''$ we find that
\begin{flalign}
	P_{\rm sm}^M(t_m,t_{\rm int})&\approx n_\gamma^2\left[\mathcal{G}_{i_M}(t_m,t_{\rm int})+\frac{\mathcal{G}_{i_M}(t_m,0)+\mathcal{G}_{i_M}(t_m+t_{\rm int},0)}{4}\right]\label{eq:gf_adjust2}
\end{flalign}
Therefore, the Green's function relevant to the multimode case, Eq. \eqref{eq:2tGf}, can be extracted from the single mode system.
We note that $\mathcal{G}_{i_M}(t_m,0)$ can be found by using a single pulse.  Thus by choosing a proper dissipative process ($\hat{Y}$) and appropriately short pulses, we can reconstruct $\mathcal{G}_{i_M}$ and simulate the matter response from the continuum mode case.
\subsection{Sampling error}
In practice $P^M_{\rm sm}$ must be estimated from multiple trials resulting in a measurement of state $i$, which will either be populated or not.  Thus the statistics governing the measured $P^M_{\rm sm}$ will be that of a binomial distribution and the sampling error in the estimate of $P^M_{\rm sm}$ is given by
\begin{flalign*}
	\sigma_{\rm sm}(t_m,t_{\rm int})=\sqrt{\frac{P^M_{\rm sm}(t_m,t_{\rm int})\left[1-P^M_{\rm sm}(t_m,t_{\rm int})\right]}{N}}
\end{flalign*}
where $N$ is the number of trials.

Then the sampling error in $\mathcal{G}_i$ is
\begin{flalign}
	\sigma_\mathcal{G}(t_m,0)&=\frac{2}{n_\gamma^2}\sqrt{\frac{P^M_{\rm sm}(t_m,0)\left[1-P^M_{\rm sm}(t_m,0)\right]}{N}}\nnl
	\sigma_\mathcal{G}(t_m,t_{\rm int})&=\sqrt{\frac{1}{n_\gamma^4}\frac{P^M_{\rm sm}(t_m,t_{\rm int})\left[1-P^M_{\rm sm}(t_m,t_{\rm int})\right]}{N}+\frac{\sigma_\mathcal{G}(t_m,0)^2+\sigma_\mathcal{G}(t_m+t_{\rm int},0)^2}{16}}.
	\label{eq:sigma_G}
\end{flalign}
Note that the impact of larger $n_\gamma$ is to reduce the sampling error estimates $\sigma_\mathcal{G}$ by an overall factor of approximately $1/n_\gamma$; this must be weighed against the error present in the underlying $P^M_{\rm sm}$ (and any additional experimental sources of error), with the optimal value being case dependent. We will return to this point later on when considering example systems.

Further, can develop an error bound on the population esimates by discretizing the integral in Eq. (\ref{eq:pop_conv}). Consider the response to a particular wavepacket with temporal profile $\varepsilon(t)$, the overall error in the simulated population is
\begin{flalign*}
	\sigma^2_P(t)&=E\left[P^M(t)^2\right]-E\left[P^M(t)\right]^2\\
			   &\approx \sum_{ijkl} \Delta t^4\varepsilon(t_i)\varepsilon(t_j)\varepsilon(t_k)\varepsilon(t_l)\big(E\left[\mathcal{G}_{i_M}(t-t_i,t_i-t_j)\mathcal{G}_{i_M}(t-t_k,t_k-t_l)\right]\\
			   &-E\left[\mathcal{G}_{i_M}(t-t_i,t_i-t_j)\right]E\left[\mathcal{G}_{i_M}(t-t_k,t_k-t_l)\right]\big)\\
			   &\approx \sum_{ijkl} \Delta t^4\varepsilon(t_i)\varepsilon(t_j)\varepsilon(t_k)\varepsilon(t_l){\rm cov}\left[\mathcal{G}_{i_M}(t-t_i,t_i-t_j),\mathcal{G}_{i_M}(t-t_k,t_k-t_l)\right]\\
			   &\approx \sum_{ij} \Delta t^4\varepsilon(t_i)^2\varepsilon(t_j)^2\sigma^2_\mathcal{G}(t-t_i,t_i-t_j),
\end{flalign*}
where $t_i$ are times spaced $\Delta t$ apart, and the first step above assumes that $\Delta t$ is small enough that the integral in Eq. (\ref{eq:pop_conv}) is well approximated by the discrete sum.  It is worth examining the case where the error in $\mathcal{G}_{i_M}(t_m,t_{\rm int})$ can be taken as roughly the same for all $t_m,t_{\rm int}$; \ie $\sigma_\mathcal{G}(t_m,t_{\rm int})\approx\sigma_\mathcal{G}$.  In that case, the normalization of the wavepacket means that
\begin{flalign}
	\sigma_P(t)\approx  \Delta t \sigma_\mathcal{G}.\label{eq:err_approx}
\end{flalign}
Since $\Delta t\propto 1/n_{\rm int}$ and $\sigma_\mathcal{G}\propto1/\sqrt{N}$, the total number of experiments needed to obtain a given error in the wavepacket response can be actually be minimized in principle by taking a larger number of intervals and rather than a larger number of samples per interval.

\subsection{Simulating dissipation}
In the previous subsection, we showed that the response of the matter system to single photon pulses could be recovered with a trapped ion simulator with a modulated ion-vibration coupling, and where a vibrational mode simulates a temporal mode of the photon field. However, to do this we had to introduce an additional Markovian dissipative process $\mathcal{D}[\hat{Y}]$ acting on the ion internal states, and set $\hat{Y}=\hat{L}$. We now examine challenges posed by this term.

As described above, additional dissipative processes can be introduced into a trapped ion platform (\eg through optical pumping), but these act locally and independently on each ion. However, the $\hat{L}$ operator is non-local operator acting on all excited states in the model and thus, in a correlated manner on several ions -- \ie if the optically coupled levels in $\hat{H}_M$ are encoded in the excited states of $N$ ions, $\hat{L}=\sum_{j}^Nl_j\hat{\sigma}^{-}_j$. This mismatch makes setting $\hat{Y}=\hat{L}$ non-trivial.
The most straightforward solution is to transform the basis so that $\hat{L}$ is local, however, operators associated with any additional decoherence processes will also be transformed, and may become nonlocal.  In the general case there will not be a basis in which all operators are local;  additionally, the state to be measured may become a superposition of several ionic states, compounding the difficulty. To overcome these difficulties, we develop a method for simulating non-local decay-type processes of the same form as $\hat{L}$ using tailored couplings to an auxiliary ion that undergoes a local, fast decay.

We introduce an auxiliary ion and isolate two internal states within it governed by a two-level system with Hamiltonian $\hat{H}_\ax=\frac{\omega_\ax}{2}(\hat{1}-\hat{\sigma}_\ax^z)$. Then we assume that the excited state is subject to a fast decay process: $\mathcal{D}[\hat{X}]$, with $\hat{X}=\chi \hat{\sigma}_\ax^-$. This could be engineered through optical pumping to a higher lying state, for example \cite{Barreiro_Blatt_2011}. In Appendix \ref{app:diss} we show that by coherently coupling the system ions to this fast decaying auxiliary ion with a Hamiltonian of the form
\begin{align}
	\hat{H}_{M-\ax} =\hat{J}\hat{\sigma}^+_\ax+h.c.=\sum_iJ_i\hat{\sigma}^-_i\hat{\sigma}^+_\ax + h.c.,
\end{align}
and adiabatically eliminating the auxiliary ion, results in a Markovian dissipative process on the system ions that takes the form $\mathcal{D}[\frac{2\hat{J}}{\chi}]$. Therefore, we can engineer the non-local decay terms necessary to model the dissipative effect of coupling to a continuum of modes by tuning the coherent couplings in $\hat{J}$ such that $\frac{2\hat{J}}{\chi} = \hat{L}$.
Thus, using a single bosonic mode and an auxiliary ion, we can simulate the response to a single-photon traversing the system in a free-field setting.

\begin{figure}
	\centering
	\subfigure[]{\includegraphics[width=.4\columnwidth]{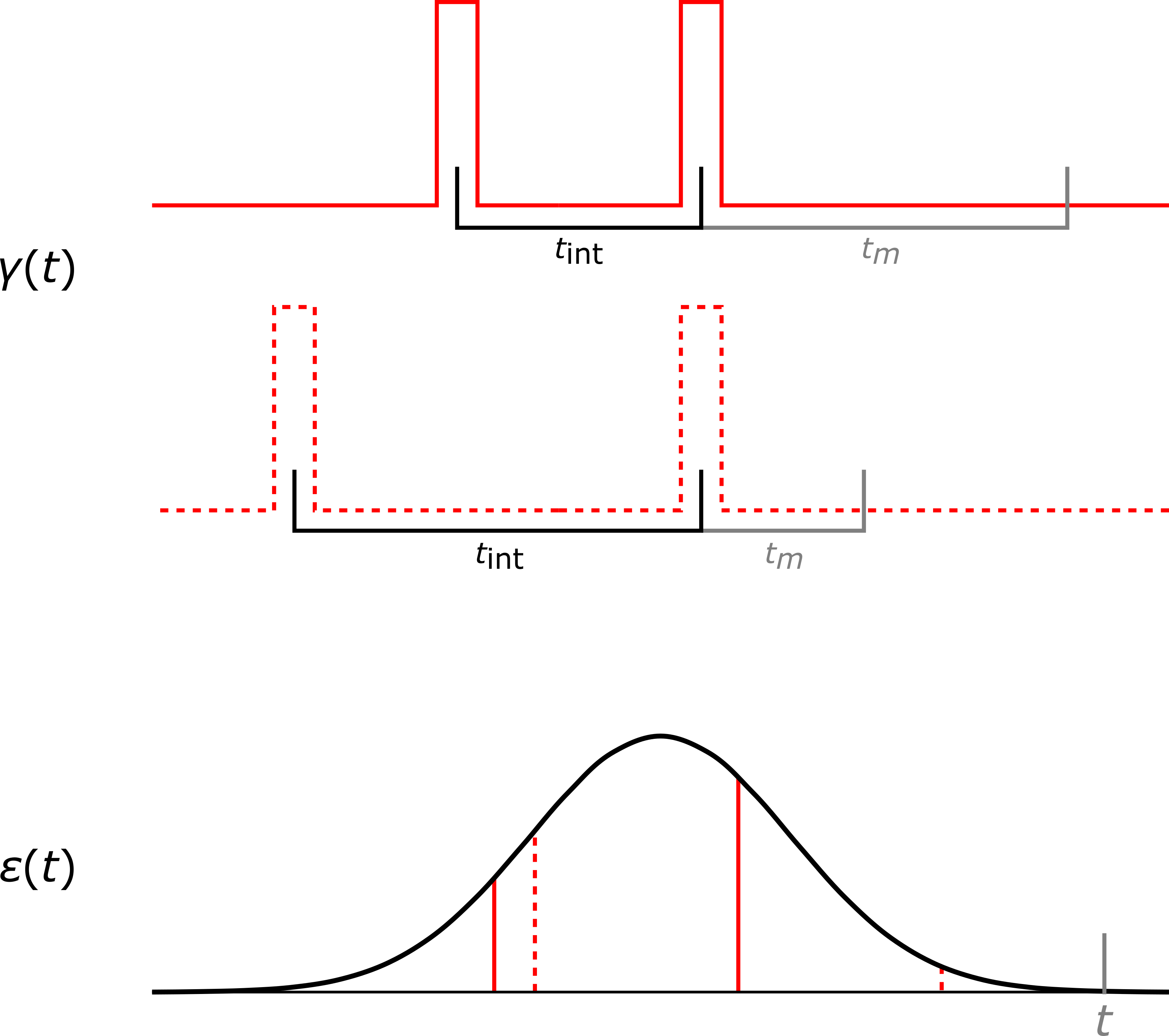}\label{fig:pulse_scheme}}
	\hspace{1cm}
	\subfigure[]{\includegraphics[width=.2\columnwidth]{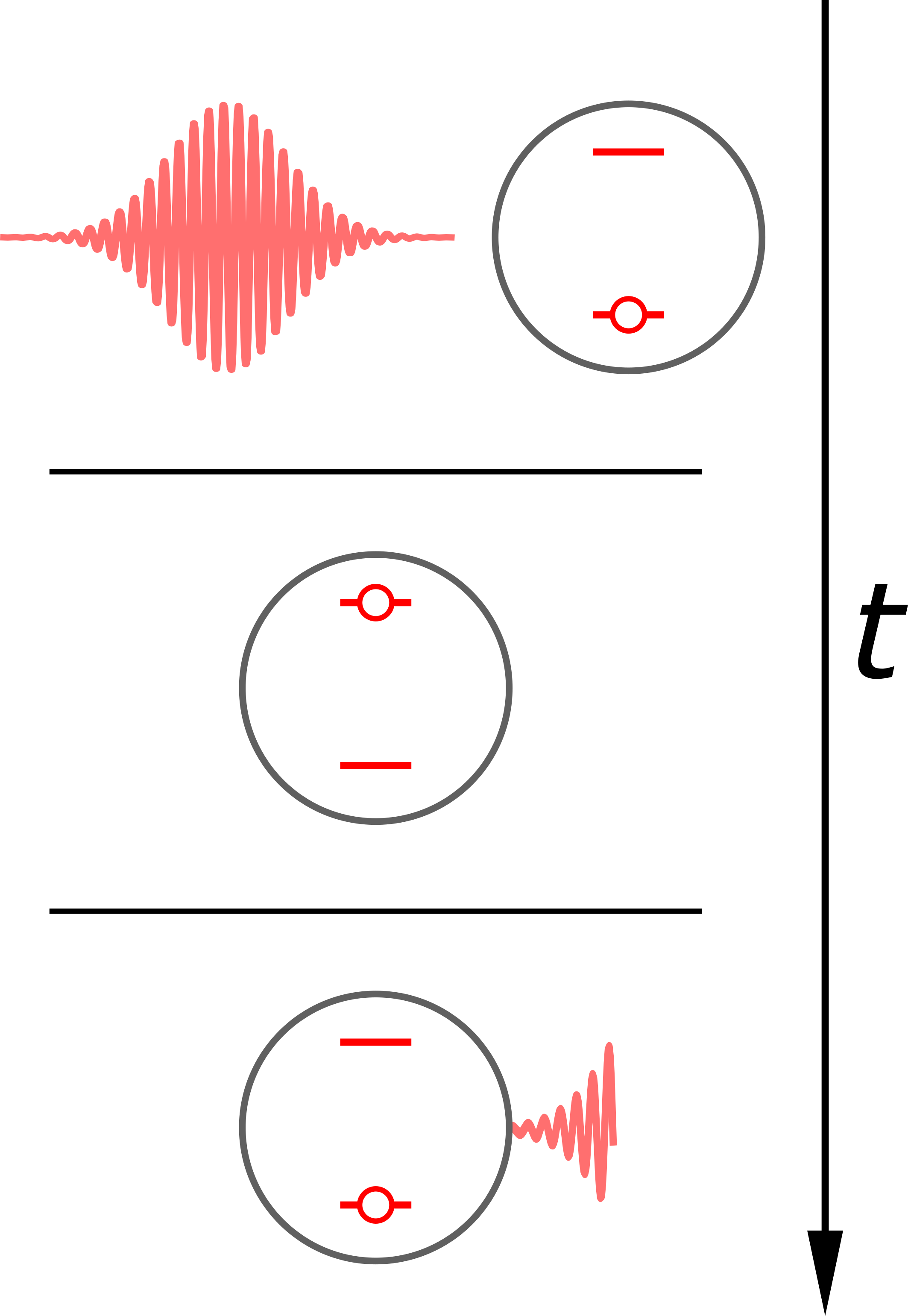}\label{fig:scheme}}
	\hspace{1cm}
	\subfigure[]{\includegraphics[width=.2\columnwidth]{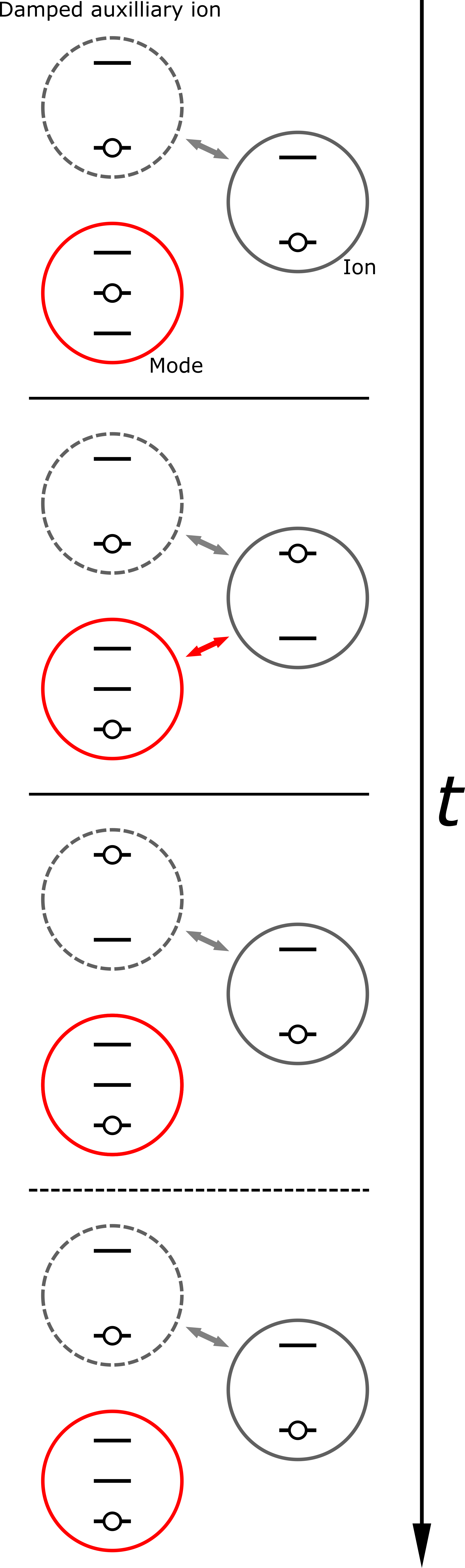}\label{fig:schemeaux}}
	\caption{\footnotesize{\bf \subref{fig:pulse_scheme}} Example pulse/measurement schedules (top) and a sketch of their inclusion in the response reconstruction (bottom).  The pulse profiles $\gamma(t)$ (red) are chosen with spacing $t_{\rm int}$ and measurements are performed at $t_m$ after the second pulse, allowing for the generation of samples of $\mathcal{G}{i_M}$ according to Eq.~\eqref{eq:gf_adjust2}.  For an arbitrary wavepacket profile $\varepsilon(t)$, the response at a time $t$ is simulated by multiplying $\mathcal{G}_{i_M}(t_m,t_{\rm int})$ by $\varepsilon(t-t_m-t_{\rm int})\varepsilon(t-t_m)$ and summing over relevant $t_m$, $t_{\rm int}$ (Eq.~\eqref{eq:pop_conv}).  {\bf \subref{fig:scheme}} A depiction of a system interacting with a field characterized by a continuum with a single-photon wavepacket (red oscillatory lineshape), traversing the system.   At early times the matter system (black circle), represented here as a two-level system with a decay mode (red lines), interacts only with the vacuum and remains in the ground state.  At some point during the traversal of the wavepacket the matter system may absorb the photon and enter the excited state, leaving the field in the vacuum state.  Later, the matter system may emit a photon back into the field, decaying to the ground state; afterwards the photon is not present at the matter system and may not be reabsorbed.  {\bf \subref{fig:schemeaux}} A system comprising a single mode for the field (red circle) whose coupling can be rapidly adjusted with time and a highly incoherent bath (dashed line circle) coherently coupled (gray double arrow) to the matter system (black circle).  Initially the boson mode is in the first excited state and uncoupled to the matter system.  At some time the interaction (red double arrow) is rapidly turned on and off, allowing for possible transfer of the excitation. At some point later this population will be transferred to the bath and almost immediately decay, mimicking a spontaneous emission event. Shown is the response for a single coupling pulse; as discussed in the text, in general multiple pulses are necessary to capture the impact of interference between excited matter states created by interaction with an external field at different times. \label{fig:schemes}}
\end{figure}

A graphical representation of the actual and simulated process is depicted in Fig.~\ref{fig:schemes}.

In practice, this engineered non-local dissipation, which is the ``desired'' relaxation dynamics, will be in competition with local, single qubit relaxation processes. For trapped ions intrinsic relaxation occurs on very long timescales and therefore $||\hat{L}||$ is likely to be a much faster rate than local relaxation rates. However, in the unlikely event that single qubit relaxation rates dominate the non-local desired relaxation prescribed by $\hat{L}$, then all simulation parameters should be scaled up such that the slowest rate is faster than single qubit relaxation and dephasing rates. 

\subsection{Experimental protocol}

Putting together the ingredients from the previous subsections, we can specify an experimental protocol to follow to measure the response functions to weak-fields and estimate the response to an arbitrary wavepacket using a quantum simulator (Fig.~\ref{fig:pulse_scheme}):
\begin{enumerate}
	\item Initialize ``matter'' system in its ground state, the auxiliary system in the ground state, and the bosonic mode in the first excited state with the coupling set to zero.
	\item Modulate coupling with the profile $\gamma(t)$, providing a ``pulse'' at time $-t_{\rm int}$.
	\item At time 0 modulate the coupling again to provide a second ``pulse''.
	\item At time $t_m$ measure the relevant population $\rho_{ii}$ of the matter system's state to obtain a sample of $P^M_{\rm sm}(t_m, t_{\rm int})$.
	\item Perform 1-4 for different $t_{\rm int},t_m$, repeat $N$ times (as needed for the desired statistical error on the estimates, see Eq. (\ref{eq:sigma_G})), to generate the Green's function $\mathcal{G}_i(t_m, t_{\rm int})$.
	\item Estimate the simulated response to a wavepacket with profile $\varepsilon(t)$ using Eq.~\eqref{eq:pop_conv}.
\end{enumerate}

This protocol is resolving the two-dimensional function $\mathcal{G}_i(t_m, t_{\rm int})$, and the total number of experiments required is a function of the resolution required in $t_m$ and $t_{\rm int}$. The maximum required resolution is set by the fastest timescale in the system evolution ($\mathcal{R} \equiv \frac{1}{\max(|| \hat{H}_M||, ||\hat{Y}||, |\gamma|)}$), which sets the required sampling resolution and discretization of $t_m$ and $t_{\rm int}$, while the optical decay rate ($|\gamma|$), sets the maximum time interval over which there is interesting dynamics. These are conservative estimates, especially the temporal resolution of $1/\mathcal{R}$, since the timescale of system evolution is typically much slower than $\mathcal{R}$. We will see an example of this in Sec. \ref{sec:eg}.

\subsection{Multiple photons}
For the case of $n$ photons, the above scheme can be easily (if tediously) generalized by considering an initial bosonic mode in its $n$th excited state with coupling pulses at $2n$ different times, leading to a $2n$-time Green's function. To isolate the desired Green's function, a similar expression to Eq. (\ref{eq:gf_adjust2}) must be used so that ``duplicate'' same-time terms can be subtracted out.  For $n=2$
\begin{flalign*}
	P_{\rm sm}(t_m,t_{\rm int},t_{\rm int}',t_{\rm int}'')&\approx n_\gamma^4\bigg[\mathcal{G}_{i_M}(t_m,t_{\rm int},t_{\rm int}',t_{\rm int}'')\\
	&+\frac{\mathcal{G}_{i_M}(t_m,0,0,0)+\mathcal{G}_{i_M}(t_m+t_{\rm int},0,0,0)+\mathcal{G}_{i_M}(t_m+t_{\rm int}+t_{\rm int}',0,0,0)}{16} \nnl
	& + \frac{\mathcal{G}_{i_M}(t_m+t_{\rm int}+t_{\rm int}'+t_{\rm int}'',0,0,0)}{16}+\dots\bigg]
\end{flalign*}
where the $\dots$ indicate the additional same-time terms (\eg $\mathcal{G}_{i_M}(t_m,0,t_{\rm int}',0)$) omitted for brevity; in general there will be $(2n)!-1$ of these.

The remaining unknowns are the settings of simulation parameters -- the shape of pulse profiles, parameters of the auxiliary system, and sampling of time intervals -- required to reach the necessary limits for desired accuracy.  We will now consider a small example system to analyze the dependence on these parameters, as well as a larger example system akin to the kind this method is intended for.

\section{Illustrations}
\label{sec:eg}

\subsection{Example 1: coupled chromophores}
\begin{figure}
	\centering
	\subfigure[Simulated]{\includegraphics[width=.4\columnwidth]{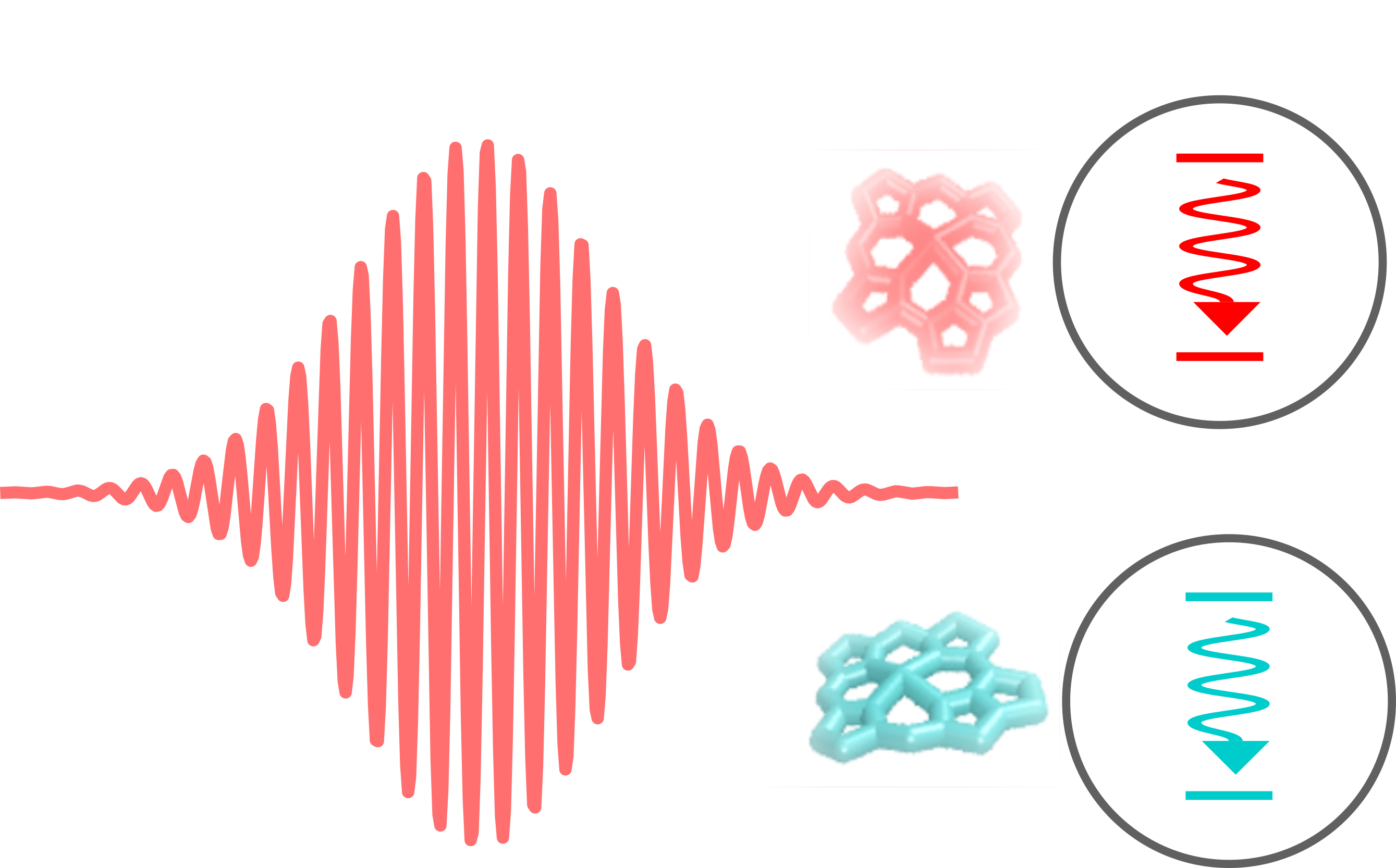}\label{fig:sim_sys}}
	\hspace{2cm}
	\subfigure[Experimental]{\includegraphics[width=.4\columnwidth]{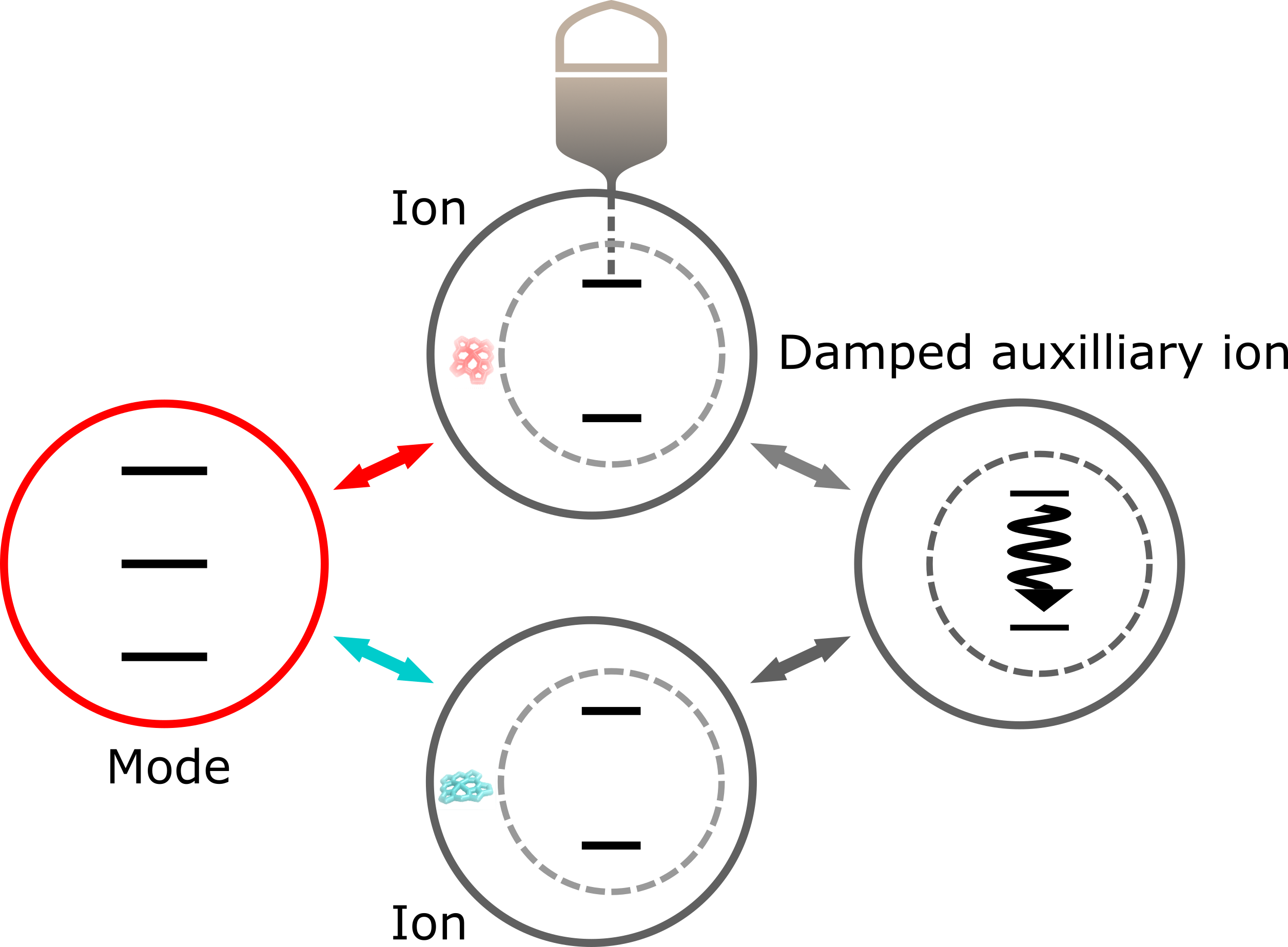}\label{fig:exp_sys}}
	\caption{Schematic for the example illustrated in Sec. \ref{sec:eg}. {\bf \subref{fig:sim_sys}} The physical scenario being simulated. Two chromophores represented by black circles are excited by a weak-field wavepacket. Each chromophore has a ground state, corresponding to a lowest unoccupied molecular orbital (LUMO), and an excited state, corresponding to the highest occupied molecular orbital (HOMO); these are drawn in red and cyan to indicate optical coupling and are shown with wavy arrows to indicate the optically mediated decay process. {\bf \subref{fig:exp_sys}} Schematic of trapped ion simulator capable of simulating the physical setup in \subref{fig:sim_sys}. The internal states of each chromophore are encoded into states of distinct ions (indicated by the dashed circles). The excited states no longer have internal decay processes, but are instead coherently coupled to an auxiliary ion that is damped. In addition, each of ions encoding chromophore states is coupled to a vibrational mode representing the field (red circle). The monitored state of the red chromophore, encoded in the top ion, is the shown attached to an output channel that represents measurement.\label{fig:systems}}
\end{figure}

To illustrate the ideas presented in the previous sections and gain intuition, we consider the simplest example of a composite matter system interacting with a weak-field wavepacket. Consider two chromophores, each with optically active excited states interacting with a single-photon Gaussian wavepacket, see Fig. \ref{fig:sim_sys}. The Hamiltonian and $\hat{L}$ operator for this example is given by:
\begin{flalign*}
	\hat{H}_M=\left[\begin{array}{ccc}
		0 & 0 & 0\\
		0 & \omega_1 & 0\\
		0 & 0 & \omega_2
	\end{array}\right], \quad \hat{L}=\left[\begin{array}{ccc}
	0 & l_1 & l_2\\
	0 & 0 & 0\\
	0 & 0 & 0
\end{array}\right],
\end{flalign*}
where the states are ordered as the common ground (HOMO) state of both molecules with zero energy, the excited (LUMO) states of molecule 1 with energy $\omega_1$, and the excited state of molecule 2 with energy $\omega_2$. Note that we have assumed there is no direct Coulomb coupling between the molecules for simplicity. We also restrict the above operators to the single excitation manifold since that is all that is needed to model the interaction with a single photon. In the multiphoton case we would need to model the multi-excitation energy states also.

Suppose we want to determine the population of state 1 in response to single photon pulses at the resonance frequency $\omega_1$.  The two-time Green's function in Eq.\eqref{eq:2tGf} for this system can be written
\begin{flalign*}\\
	\mathcal{G}_{1}(t_m,t_{\rm int})&=\sbra{1}{1}\sop{\hat{g}\dg(t_m)}{\hat{g}(t_m)}\left(\sop{\expn{i\omega_1 t_{\rm int}}\hat{g}\dg(t_{\rm int})}{\hat{1}}+\sop{\hat{1}}{\expn{-i\omega_1 t_{\rm int}}\hat{g}(t_{\rm int})}\right)\sop{\hat{L}\dg}{\hat{L}\dg}\sket{0}{0},
\end{flalign*}
with
\begin{align}
		\hat{g}(t)&=\exp\left(i\left[\begin{array}{cc}
			\omega_1 +il_1^2& +il_1l_2\\
			+il_2l_1 & \omega_2+il_2^2
		\end{array}\right]t \right).
\end{align}

We construct a trapped ion simulation of this two chromophore system with two ions, each with the following Hamiltonian and $\hat{L}$ operators:
\begin{flalign*}
	\hat{H}_1&=\left[\begin{array}{cc}
		0 & 0\\
		0 & \omega_1
	\end{array}\right]\otimes \hat{1},\quad
\hat{L}_1=\left[\begin{array}{cc}
	0 & l_1\\
	0 & 0
\end{array}\right] \otimes \hat{1}\\
	\hat{H}_2&=\hat{1}\otimes \left[\begin{array}{cc}
	0 & 0\\
	0 & \omega_2
\end{array}\right],\quad
\hat{L}_2=\hat{1}\otimes \left[\begin{array}{cc}
	0 & l_2\\
	0 & 0
\end{array}\right]\\
\hat{H}_M&=\hat{H}_1 + \hat{H}_2\\
\hat{L}&=\hat{L}_1 + \hat{L}_2
\end{flalign*}
and bosonic mode and auxiliary qubit for the field are described by
\begin{flalign*}
	\hat{H}_F&=\omega_1\hat{a}\dg\hat{a}\\
	\hat{H}_\ax &=\left[\begin{array}{cc}
		0 & 0\\
		0 & \omega_1
	\end{array}\right].
\end{flalign*}
The couplings are
\begin{flalign*}
	\hat{H}^{\rm sm}_{M-F}=\hat{L}\hat{a}\dg+h.c.\\
	\hat{H}_{M-\ax}=\frac{\chi}{2}\hat{L}\hat{\sigma}_{\rm aux}^+ + h.c.\\
\end{flalign*}
\begin{figure}
	\centering
	\subfigure[]{\includegraphics[width=.45\columnwidth]{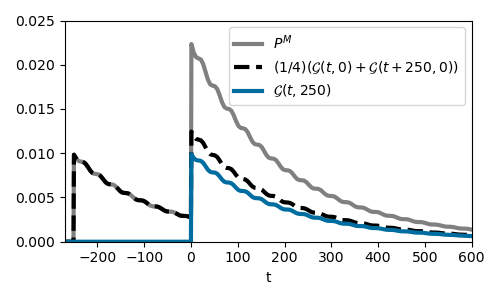}\label{fig:Gcomp}}
	\subfigure[]{\includegraphics[width=.45\columnwidth]{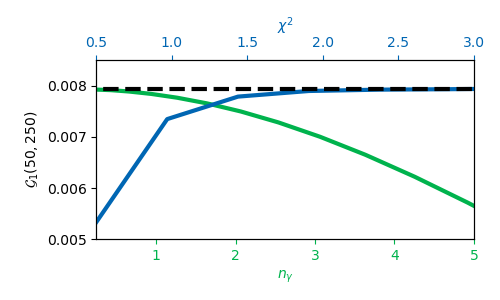}\label{fig:n_chi}}
	\caption{{\bf \subref{fig:Gcomp}} The population of the excited state of 1 in response to two $\delta$-pulses spaced $t_{\rm int}=250$ apart (gray) along with the independent contributions from each pulse (dashed line) and $\mathcal{G}_{1}(t,250)$ (blue), the latter of which captures the interference between coherent excitations due to the pulses.  {\bf \subref{fig:n_chi}} shows the impact of $n_\gamma$ and $\chi^2$ on the simulated value of $\mathcal{G}_{1}(50,250)$ in the limit of $\delta$-pulses ($t_\gamma=1$).  The dashed line marks the exact value. $\chi^2=5$ for the $n_\gamma$ sweep, and $n_\gamma=0.25$ for the $\chi^2$ sweep. \label{fig:params}}
\end{figure}

There are three parameters that must be adjusted to ensure that the simulation operates in the appropriate limit for simulating $\mathcal{G}_{1}$: $n_\gamma$,$\chi$, and $t_\gamma$.  Additionally, the times $t_m$ and $t_{\rm int}$ for which $\mathcal{G}_1$ are simulated must be chosen from a sufficiently dense grid to ensure the pulse profiles of interest can be accurately reconstructed.  In Fig.~\ref{fig:params} we show the dynamics for exact and simulation systems, given $\omega_1=1.0$, $l_1^2=0.0036$, $\omega_2=0.8$, and $l_2^2=0.0064$, as well as the impact of $n_\gamma$, and $\chi$ on the population response to two pulses used to generate $\mathcal{G}_{1}(t_m,250)$.  Fig.~\ref{fig:Gcomp} shows the exact result for a single $t_{\rm int}$. The main features are the step changes at the pulse times, the long decay due to spontaneous emission, and oscillations at frequency $\omega_1-\omega_2$ due to the different energies of states 1 and 2.  We emphasize that, as given by Eq.~\eqref{eq:gf_adjust2}, in order to obtain $\mathcal{G}_{1}(t,250)$ we must subtract out the $t_{\rm int}=0$ from the overall response to the two pulses. Shown in Fig.~\ref{fig:n_chi}, in this case the necessary parameter values required to reach the asymptotic limit.  $\chi_2$ must essentially be about two to three orders of magnitude faster than the rates $l_i^2$ (which are relatively small).
To understand the impact of $n_\gamma$, recall that this parameter corresponds to the area of the pulses of coupling between the ions and the mode. Therefore, larger $n_\gamma$ means greater population exchange between the mode and internal states of the ion, and thus increased reabsorption. This explains the deviation of the computed response and true response with increasing $n_\gamma$. $n_\gamma \sim 1$ (resulting in the expected population shown in Fig.~\ref{fig:Gcomp}) is sufficient for most purposes.

\begin{figure}
	\centering
	\subfigure[]{\includegraphics[width=.45\columnwidth]{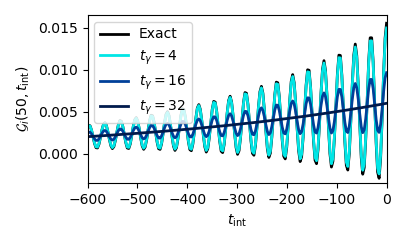}\label{fig:g2}}
	\subfigure[]{\includegraphics[width=.45\columnwidth]{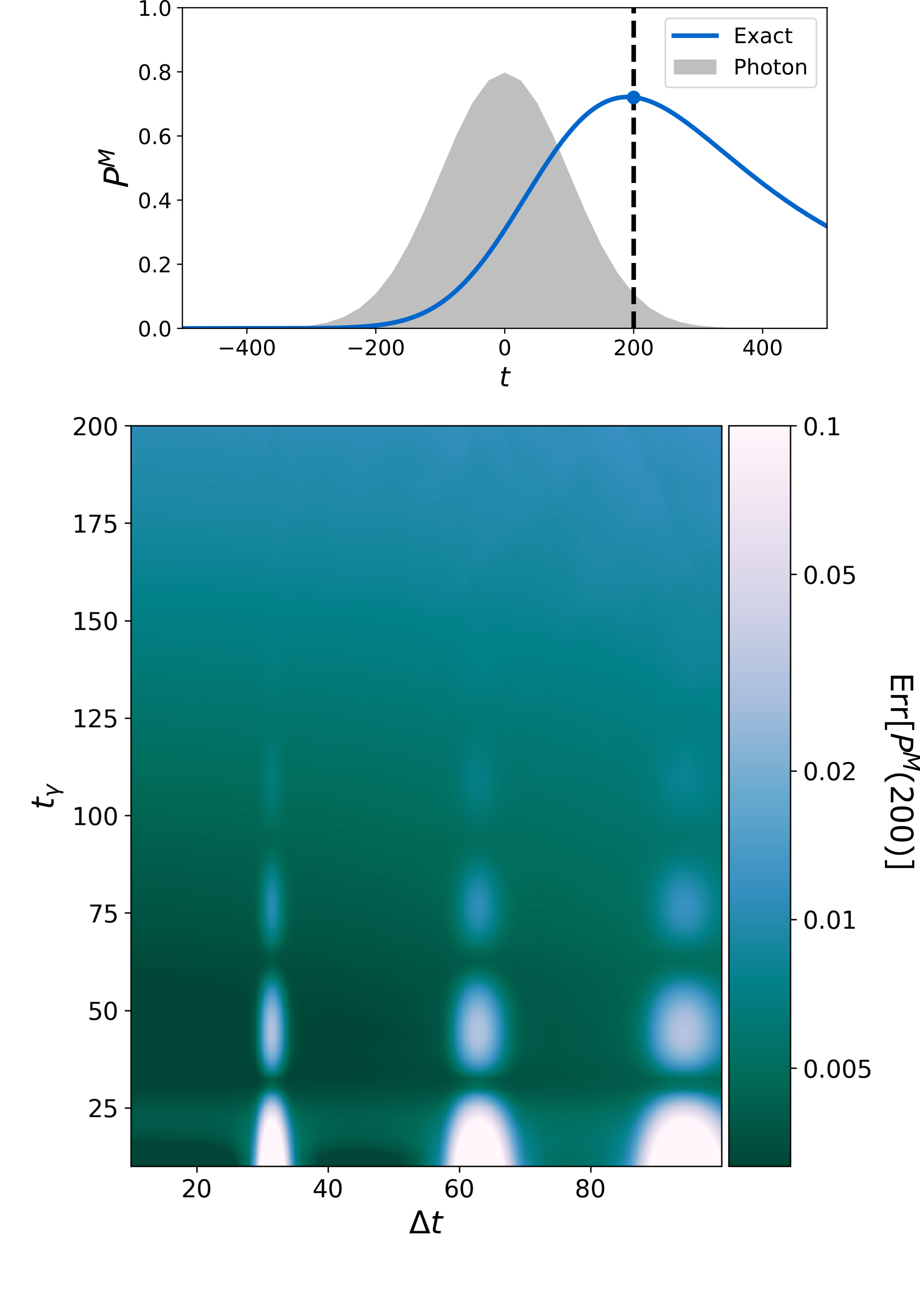}\label{fig:t2d}}

	\caption{{\bf \subref{fig:g2}} A plot of $\mathcal{G}_{1}(200,t)$ for different $t_\gamma$.  {\bf \subref{fig:t2d}} The population of excited state 1 in response to a single photon with a Gaussian-shaped temporal profile. The top panel shows the exact response $P^M(t)$ with the simulated time of measurement marked by a vertical dashed line.  The gray shaded area shows the profile shape of the intensity of the photon pulse. The contour plot in the bottom panel shows the relationship between the error (computed population less the exact value indicated in the top panel) in the simulated population  at $t=200$ for different values of $t_\gamma$ and sampling intervals $\Delta t$ of $t_m$ and $t_{\rm int}$, for $n_\gamma=1$ and $\chi^2=5$.   \label{fig:t_int}}
\end{figure}
The impact of different choices of $t_\gamma$ is more nuanced.  While a larger $t_\gamma$, which results in wider pulses, might seem to result in a poor approximation of a delta impulse (which is what the pulse mimics), the fact that the Green's function is convolved with a smooth wavepacket profile results in considerable tolerance to the value of $t_\gamma$. To demonstrate this, consider Fig.~\ref{fig:g2}, which shows the $\mathcal{G}_{1}$ computed for different pulse widths.  For narrow pulses, oscillations due to the difference $\omega_1-\omega_2$ are quite pronounced, while for wider pulses these are substantially reduced.  However, since according to Eq.~(\ref{eq:2tGf}) these computed Green's function are further convolved with the wavepacket profile, if the wavepacket is slowly varying in time the impact on the final result of larger $t_\gamma$ is minimal until it begins to exceed the wavepacket's temporal width.  This is demonstrated in Fig.~\ref{fig:t2d} for a Gaussian wavepacket with $\sigma_t=100$.  The inset shows the exact response $P^M(t)$, while the contour plot shows the deviation from the exact value for $P^M(200)$ for different values of $t_\gamma$ and sampling interval $\Delta t$ of $t_m$ and $t_{\rm int}$. The computed response is robust to fairly large $t_\gamma$; in fact, larger values are somewhat preferable if one wishes to minimize the impact of choice of the $\Delta t$ sampling interval, which exhibits pathologies at particular values.  These are at multiples of $2\pi/(\omega_1-\omega_2)=31.412$ and are due aliasing from sampling intervals that align with the period of oscillation of $\mathcal{G}_{1}$.

Note that the timescale of system evolution in this example is $2\pi/(\omega_1-\omega_2)$, which is significantly larger than $ \frac{1}{||H_{M}||}$ since the coherent dynamics is only between the excited states in the system, and the optical decay rate is much slower than the coherent timescales. This means that the resolution required of the two-dimensional function $\mathcal{G}_1(t_m, t_{\rm int})$ is na\"ively $\Delta t <  2\pi/(\omega_1-\omega_2)$, and indeed, the error in Fig. \ref{fig:t_int}(b) (for constant $t_\gamma$) is smallest for $\Delta t$ in this range. However, note that the error is also small for larger $\Delta t$ values as long as the aforementioned pathological values are avoided. This small error at larger $\Delta t$ values results from a cancellation of errors due to the integral in Eq. (\ref{eq:pop_conv}); \ie the fast oscillations of $\mathcal{G}_1$ and slow variation of $\epsilon(t)$ imply that the integral evaluates to almost zero, even if $\mathcal{G}_1$ is coarsely sampled. This is specific to this system, and not generically expected.

\begin{figure}
	\centering
	\subfigure[]{\includegraphics[width=.5\columnwidth]{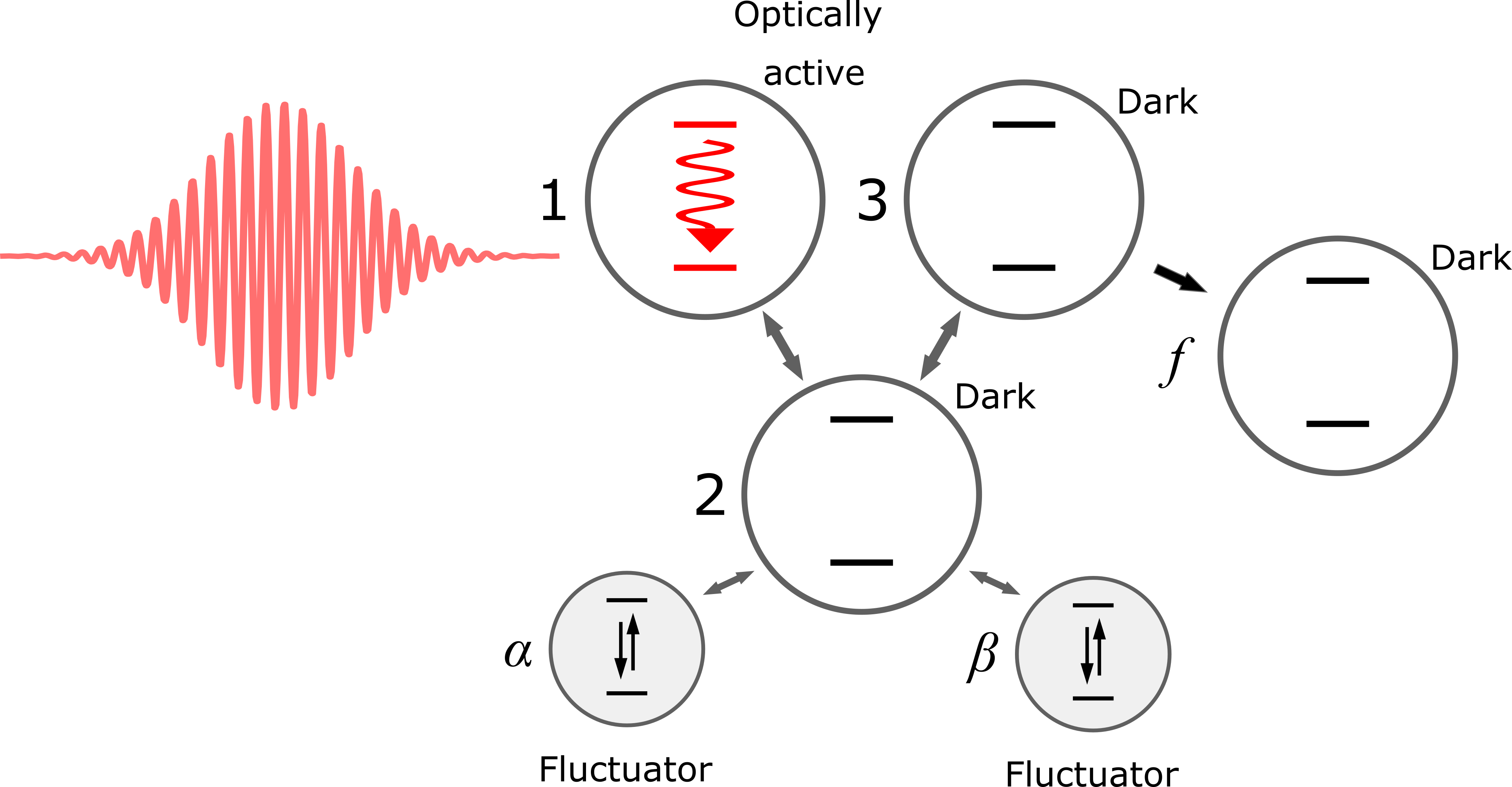}\label{fig:large}}\quad\quad\quad
	\subfigure[]{\includegraphics[width=.35\columnwidth]{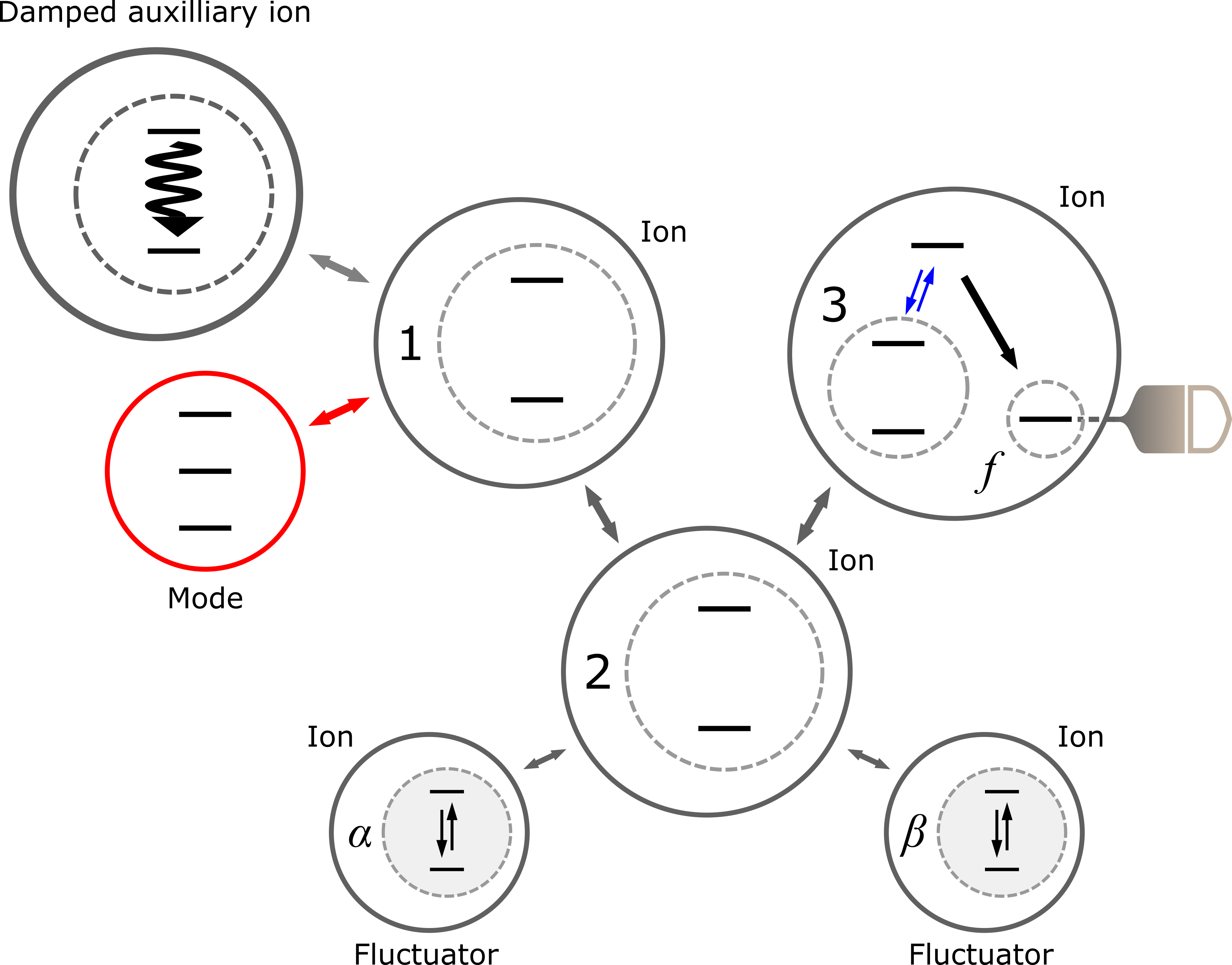}\label{fig:largeaux}}
	\subfigure[]{\includegraphics[width=.45\columnwidth]{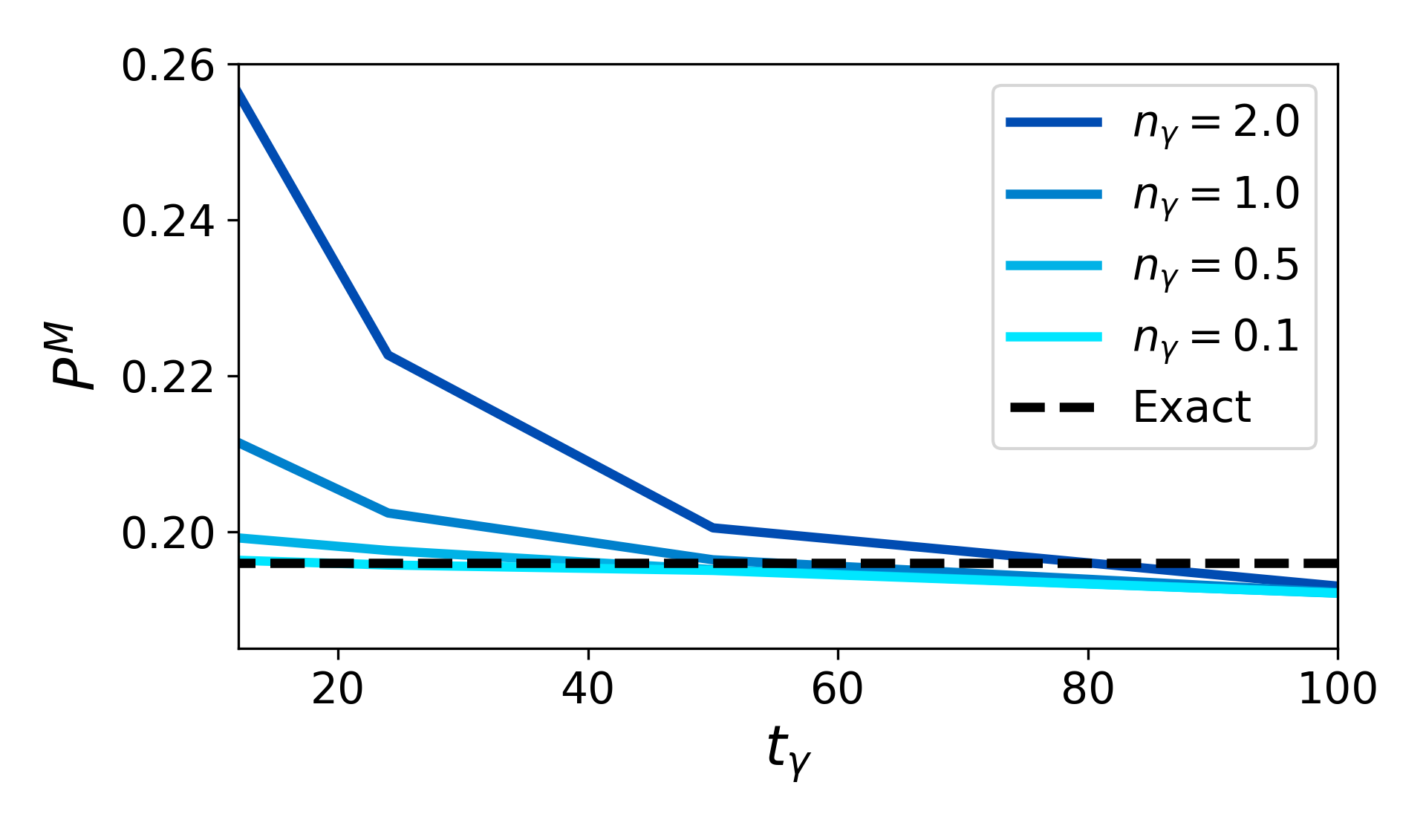}\label{fig:largecomp}}
	\subfigure[]{\includegraphics[width=.45\columnwidth]{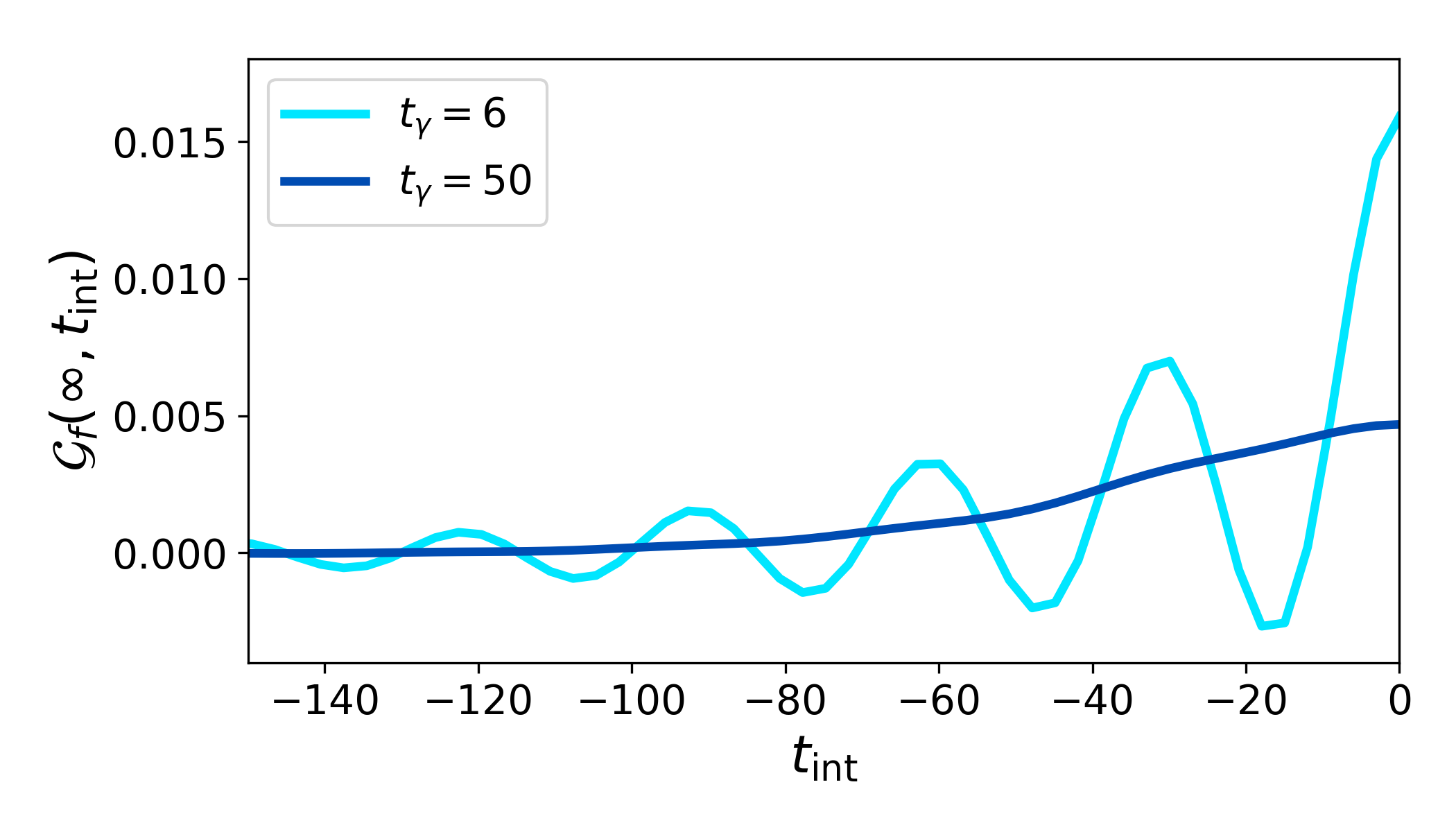}\label{fig:g2_tg_large}}
	\caption{{\bf \subref{fig:large}} Schematic of the Example 2. This model has one optically active element, \eg a quantum dot, that couples to a chain of optically inactive (dark) elements, which could be other quantum dots. We are interested in the transfer of a photoexcitation down the chain to a state ($f$) in the final dark element. Note that the two-headed arrows indicate coherent coupling between elements and the single-headed arrow indicate incoherent transfer. Making this transfer difficult to model is that one of the intermediate dark elements (2) is coupled to classical fluctuators (shaded circles), which induce non-Markovian stochastic dynamics in the remainder of the system. {\bf \subref{fig:largeaux}} The trapped ion simulation is setup similarly to the small system. Again, we model the internal states of each of the quantum dots with internal states of ions (shown in dashed circles). The coherent coupling between quantum dots and fluctuators are implemented through Hamiltonian engineering (see main text). Ion 3 encodes both states in quantum dot 3 and also the $f$ state. The incoherent coupling to state $f$ is implemented through optical pumping via an intermediate state (similar to ion readout).  {\bf\subref{fig:largecomp}} The measured population of the end state for different $n_\gamma$ as a function of $t_\gamma=2\Delta t_{\rm int}$.  {\bf \subref{fig:g2_tg_large}} The $\mathcal{G}_{f}(\infty,t)$ that describes the response of the system to input pulses of varying width. In both cases, $\omega_0=\omega_1=1.0$, $\omega_2=0.8$, $\omega_3=0.9$,$\omega_f=0.5$,$\omega_\alpha=\omega_\beta=0.1$,$J_{12}=0.096$,$J_{23}=0.1$,$J_{2\alpha}=J_{2\beta}=0.02$, $l=0.01$, $\Gamma_f^2=0.4$, $\Gamma_\alpha^2=0.25$,$\Gamma_\beta^2=0.5$, $n_\gamma=1$ and $\chi^2=5$.   \label{fig:large_systems}}
\end{figure}

\subsection{Example 2: energy transfer with non-Markovian environment}

The system considered in the last section is of course trivial to simulate conventionally.  Even in more complicated cases, single photon response can often be efficiently simulated by restricting the calculation to an energetically accessible subspace.  However, this is not always especially helpful, as in many systems -- especially those featuring many body interactions -- the single (or few) photon accessible manifold can nonetheless be quite large and involve subsystems where multiple energy states must necessarily be included.  For example, an important consideration and phenomenon of interest is the dynamics and behavior of such systems under the influence of non-Markovian (\eg $1/f$) noise due to impurities such as charge traps at interfaces. These can be modeled by adding auxiliary -- frequently two-level -- systems driven by Markovian noise coupled to the main system of interest. Scaling due to these additional degrees of freedom added by considering explicitly structured noise sources cannot be reduced, quickly rendering conventional simulation intractable.  The present scheme, however, is not impacted by such considerations; as long as the necessary elements can be incorporated into the experimental setup, the number of samples required experiences constant scaling.

To illustrate this, we now consider a model that is representative of the kind of complex material whose response one might want to calculate with our approach. We model an optically active molecule or quantum dot that is Coulomb coupled to several optically inactive molecules/dots. The system is also coupled to environmental fluctuators that cause non-Markovian decoherence of the system. We are interested in the energy transfer dynamics induced by single photon illumination in such a system and thus monitor the population at a optically inactive site. Due to the non-Markovian dynamics, simple reduced treatments of the optically activated dynamics are not possible. This minimal model is representative of the structures found in biochemical molecular complexes or quantum dot nanostructures. In Fig. \ref{fig:large_systems}(a) we depict such a system: an optically active subsystem on the left, coherently coupled to two dark systems with an incoherent decay to a stable state at the end, mimicking an electron transport pathway.  The middle state of the coherent chain is coupled to a pair two-level fluctuators acting as a non-Markovian noise source.

The matter Hamiltonian for this example is given by
\begin{flalign}
	\hat{H}_M&=\sum_i\hat{H}_i+\sum_{\langle i,j \rangle}\hat{H}_{ij}+\sum_j\hat{H}_{2j} \label{eq:eg2_ham}\\
	\hat{H}_i&=\omega_i\hat{\sigma}^+_i\hat{\sigma}^-_i,\quad i\in 1,2,3,f,\alpha,\beta\\
	\hat{H}_{ij}&=J_{ij}\hat{\sigma}^+_i\hat{\sigma}^-_j,\quad i,j\in 1,2,3 \label{eq:couplings}\\
	\hat{H}_{2j}&=J_{2j}\hat{\sigma}^+_2\hat{\sigma}^-_2\hat{\sigma}^+_j\hat{\sigma}^-_j,\quad j \label{eq:dot-fluct} \in \alpha,\beta
\end{flalign}
where $1,2,3,f$ index the quantum dot subsystems and $\alpha$ and $\beta$ are the two fluctuators. $\hat{H}_i$ represent local energies of the quantum dot excited states and fluctuator energies, $\hat{H}_{ij}$ represent the coherent Coulomb coupling between the quantum dots with $\langle i,j\rangle$ indexing the dots that are coupled (see double headed arrows in Fig. \ref{fig:large_systems}(a)), and $\hat{H}_{2j}$ representing the coupling between the quantum dot excited states and fluctuator states.

The field coupling is only to the bright dot and therefore,
\begin{flalign}
	\hat{L}=l\hat{\sigma}^-_1.
	\label{eq:eg2_L}
\end{flalign}
Additionally, the system is driven by two incoherent processes, generated by Lindblad terms $\mathcal{D}[\hat{Z}]$, with the following generators
\begin{flalign}
	\hat{Z}_f&=\Gamma_f\hat{\sigma}^-_3\hat{\sigma}^+_f \label{eq:zf} \\
	\hat{Z}_i&=\Gamma_i\hat{\sigma}^x_i, \quad i\in \alpha,\beta. \label{eq:zi}
\end{flalign}
The first of these represents the incoherent transfer from the third (dark) dot to a final stable state $f$ on the last dot in the chain (which is monitored by the electron transfer pathway/measurement), and the second represents the incoherent flipping of the fluctuator state. The reduced dynamics of the quantum dots is non-Markovian and complex due to the coupling to the fluctuators and the EM field.

This system is simulated with the trapped-ion platform using the scheme shown in Fig. \ref{fig:large_systems}(b). Individual ions represent the active and dark quantum dots and the fluctuators. In addition, a vibrational mode represents the EM field and we introduce an auxiliary ion to model the field damping as described above. The terms in the matter Hamiltonian $\hat{H}_M$ are simulated using either local Stark shifts to ion energies or through M\o lmer-S\o renson interactions in the rotating wave regime (for the coherent interactions in $\hat{H}_{ij}$ (Eq. \eqref{eq:couplings}). The coupling between the quantum dot $2$ and the fluctuators, $\hat{H}_{2j}$ in Eq. \eqref{eq:dot-fluct}, is not of any of the standard forms discussed in Sec. \ref{sec:ions}. It describes a shift in energy of ion 2, that depends on the state of the fluctuators. This type of interaction, although not as popular in trapped-ion quantum computing as the M\o lmer-S\o renson interaction, can be engineered and has been demonstrated \cite{Leibfried_2003,Ballance_2016}. Therefore, this example also requires such $\sigma^z\otimes \sigma^z$ interactions between ion 2 and the ions modeling the fluctuators.

The incoherent dynamics generated by Eq. \eqref{eq:zi} are generated by driving each of the fluctuator ions with microwave fields superimposed with broadband noise \cite{Potocnik_2018}. Finally, the incoherent coupling to the stable state $f$, generated by Eq. \eqref{eq:zf}, is achieved by encoding this state as a metastable hyperfine state of ion 3. Then, by optically pumping the excited state of ion 3 to a high-lying state that decays into the state f, we achieve incoherent coupling desired. The strength of the coupling can be tuned by the detuning of the optical pump.

As with Example 1, the spontaneous emission (decay) from the optically active quantum dot is modeled through a coupling to an auxiliary ion that is damped. The coupling between the ion 1 and this auxiliary ion, and the vibrational mode that models the optical field, and the internal Hamiltonians for both of these systems take the form:
\begin{flalign*}
                \hat{H}_F&=\omega_1\hat{a}\dg\hat{a}\\
                \hat{H}_\ax &=\left[\begin{array}{cc}
                                0 & 0\\
                                0 & \omega_1
                \end{array}\right]\\
                \hat{H}^{\rm sm}_{M-F}&=\hat{L}\hat{a}\dg+h.c.\\
                \hat{H}_{M-\ax}&=\frac{\chi}{2}\hat{L}\hat{\sigma}^+_\ax+h.c.,
\end{flalign*}
where $\hat{L}$, as in Eq. \eqref{eq:eg2_L}, is a local operator acting on ion 1. As described in previous sections, the auxiliary ion is optically pumped in order to simulate dissipation.

In Fig.~\ref{fig:large_systems}(c,d) we show the response characteristics for a measurement of the stable state $f$ following the passage of a Gaussian photon pulse of the same kind considered above.  In Fig.~\ref{fig:large_systems}(d) the function $\mathcal{G}_{f}(\infty,t_{\rm int})$ is plotted. The decay processes result in a sharper response with respect to the interval, reducing the overall window over which $\mathcal{G}_{f}$ is significant and must be computed. In Fig.~\ref{fig:large_systems}(c) we show the expected population after the simulated pulse has passed and the population of $f$ has stabilized for varying $t_\gamma$ and $n_\gamma$.  As before, the final result is robust for rather large choices of $t_\gamma$ and $\Delta t_{\rm int}$.  Interestingly, the longer pulse widths act to mitigate this error; by spreading out the excitation over a longer period, the incoherent process are given more time to damp out the excitation, limiting accumulation in the 1 state that can be coherently transferred back to the bosonic mode, resulting in less error as $n_\gamma$ is increased.

Finally, we consider the number of trials $N$ necessary to estimate $\mathcal{G}$.  For both cases the values of $P^M_{\rm sm}$ are around 0.01.  This suggests that, in order to obtain $\sigma_\mathcal{G}<0.001$ one should take $N\sim10^4$ samples. If we consider a sampling interval for $t_m,t_{\rm int}$ of $\Delta t=10$ for a total time domain of 1000, then according to Eq.~\eqref{eq:err_approx} an error of 0.01 in the population corresponds to around one million experiments. As mentioned previously, depending on the experiment objectives and the system in question, there is some freedom to ``reallocate'' trials from the sampling of $\mathcal{G}$ to the sampling of $t_m$ and $t_{\rm int}$, reducing the overall number of experimental runs needed.

\section{Scaling considerations}
\label{sec:scaling}
Now we address the scalability of the proposed approach to quantum simulation of weak-field light-matter interactions. Ultimately, the goal is to simulate interactions with material systems with hundreds or thousands of localized degrees of freedom (e.g., atoms). This corresponds to hundred or thousands of qubits since each localized degree of freedom requires at least one qubit to model. Furthermore, one requires a constant number of auxiliary qubits (or other degrees-of-freedom) to capture the effects of spontaneous emission. While it is an engineering and technical challenge to scale the trapped-ion (or any other quantum computing) platform to such sizes, it is within the roadmap of the technology. The aspect of the simulation protocol that requires more thought is the scalability of engineering all the interactions necessary for the general model, given by the Hamiltonian (in the rotating frame with respect to mode $\hat{a}$'s frequency $\omega_0$):
\begin{align}
	\hat{H} &= \hat{H}_M +\hat{H}_{M-F} + H_{M-\ax}, \quad\quad \textrm{with} \nn \\
	\hat{H}_M  &= \sum_{j=1}^n \omega_0^j \hat{\sigma}^z_j + \sum_{\langle i,j \rangle} J_{ij} \hat{\sigma}^x_i \hat{\sigma}^x_j, \nn \\
	\hat{H}_{M-F}&=i \gamma(t) (\expn{i\omega_0t}\hat{L} \hat{a}\dg-\expn{-i\omega_0t}\hat{L}\dg \hat{a}) \nn\\
	\hat{H}_{M-\ax} &= \frac{\chi}{2}(L \hat{\sigma}^+_\ax + L\dg \hat{\sigma}^-_\ax),
\end{align}
where we have assumed the matter Hamiltonian $H_M$ to be of the form that can be engineered in trapped ions (\ie $H_a$ in Eq. \eqref{eq:ha}). In addition to this Hamiltonian, fast dissipation at rate $\chi^2$ must be engineered on the auxiliary qubits.

In the following, we discuss the primary concerns when implementing this model at scale on the trapped ion platform.
\begin{enumerate}
\item We begin with the need to engineer a fast decay process on one or several auxiliary ions. As mentioned above, this can be accomplished by optical pumping of an ion, i.e. driving the ions coherently with laser light followed by a spontaneous decay. The spontaneous decay process, however, generates a recoil, i.e. there is a probability of $\eta^2_\alpha(n_\alpha+1)$   that a phonon in mode $\alpha$ is generated. While this probability is small for small motional mode occupations $n_\alpha$ and typcial Lamb-Dicke parameters $\eta_\alpha$ of a few percent. Still if a large number of spontaneous emission process needs to be simulated, this heating process needs to be countered. Probably the most promising avenue to accomplish this is to employ another ion species or isotope which allows cooling using light not resonant with any of the transitions of the primary ion species. Thus, the temperature of the ion crystal can be controlled while preserving the coherence of the ions.

\item As mentioned in Sec. \ref{sec:ions}, one also must be careful about accounting for the number of motional modes required. Not only do we need a mode for the electromagnetic field (with annihilation operator $\hat{a}$), but each qubit-qubit interaction must be mediated by a separate motional mode. While the most general model can contain $n(n-1)/2$ non-zero $J_{ij}$ coupling values, in physically relevant models, the coupling structure is more sparse. For example, if these couplings arise from screened Coulomb interactions, as is common in biochemical models, the coupling values decay rapidly with separation distance and consequently, in a complex of many constituents (\emph{e.g.,} molecules) only the couplings between the closest ones have to be taken into account to get an accurate picture of the dynamics. Suppose there are $m<n(n-1)/2$ such non-zero coupling terms.
Finally, implementing all the qubit-qubit interactions in $H_{M-\ax}$ requires $p$ modes, where $p$ is the number of optically active subsystems in the model (the number of subsystems/qubits that enter the definition of the optical coupling $\hat{L}$). Therefore, in sum, we need $m+p+1$ motional modes to implement this model. A chain of $n$ ions has $3n$ total vibrational modes, and the allocation of these modes to mediate the interactions in a model will be a complex optimization that depends on  hardware and trap constraints and the model. However, the minimum necessary condition that we need to satisfy is $m+p+1 \lessapprox 3n$.
\item If the number of motional modes required by the model is feasible, the next concern is the complexity of having all interactions turned on at once to implement the model. Even assuming the technical challenges of having so many addressing laser beams can be met, one could be worried about interference or crosstalk effects between terms. Not only do many ion-mode interactions have to be on at once, but the same ion needs to interact with multiple modes at once. The main concern is that light coupling a particular ion to a particular mode with strength $\kappa$ will also couple this very same ion off-resonantly to other motional modes $\alpha$ detuned by $\Delta_\alpha$ with strength $n_{\alpha}(\kappa/\Delta_\alpha)^2$. Anticipating a typical coupling strength of order kHz and aiming at a coupling of less than 1\,\% at nearby modes, we require $\Delta_\alpha \gg n_{\alpha} \times 10\,$kHz. In order to mitigate crosstalk from strongly coupled modes to weakly coupled ones, it will be advisable to group them such that weakly coupled modes have large frequency differences to the strongly coupled ones. 
Typical motional frequencies of the transverse modes of an ion strings are of order 5\,MHz spanning about 1\,MHz. Shaping the axial potential such that those mode frequencies are uniformly spread, there is room to control the coupling to close to 100 modes. If less crosstalk should be desired or the ion crystal is not near the motional ground state, all interaction strengths could be reduced to slow down the simulation at the expense becoming more sensitive to decoherence such as motional heating and qubit decoherence, both of which can be larger than 100\,ms.
\end{enumerate}

The ultimate path forward to scaling this approach might be to go beyond analog simulation and instead use a digitized model that performs Trotterization of the above model. This would enable one to implement each interaction between subsystems/qubits separately, thereby allowing one to recycle the modes and thus reducing the required number.
Although digital simulation offers many advantages, especially for scaling up the approach, it is important to realize that digitized evolution can take longer, and thus require a platform with better operation fidelities and coherence times.

\section{Applying the response function approach to other physical settings}
\label{sec:other}
The utility of the described scheme is not limited to simulation of light-matter interactions.  First we note that the scheme can be straightforwardly applied to the study of interaction with phononic wavepackets, as the quantum mechanical description is almost identical.  A fully quantum mechanical description of phonon wavepackets and interactions is increasingly necessary as novel technologies exploiting phonons for quantum information processes emerge \cite{Tian_2010,Stannigel_2012,Wigger_2015,Soh_2021}.
Second, the wavepacket aspect is not essential; the Green's function approach and ability to treat non-locally interacting baths is generally useful for simulating and analyzing open quantum systems in regimes where statistical treatments are insufficient.  In particular, the $2n$-time Green's function described here contains information about the dynamics of coherence in the $n$th-excited state manifolds, and can be used to probe the behavior of systems coupled to complex baths that are difficult to simulate conventionally. Our scheme can be adapted to simulate the response of a many-body system to mesoscopic, structured environments with long-lived coherent, bosonic degrees of freedom. Such mesoscopic environments are increasingly common in engineered nanostructures \cite{Dutt_2006, Das_2010}.

\section{Conclusion}
\label{sec:disc}
We have presented an approach to simulating light-matter interactions using analog quantum simulators with access to controllable bosonic modes. Our approach relies on extraction of time-dependent response functions through dynamic modulation of the coupling between the bosonic mode(s) and other degrees of freedom modeling the matter subsystem. We analyzed through calculations and numerical examples the parameter regimes in which the quantum simulations produces accurate predictions. As demonstrated through examples in Sec. \ref{sec:eg}, our simulation scheme is fairly robust to choice of experimental parameters -- although we are computing Green's functions, because these quantities are convolved with smooth wavepacket profiles the represent the electromagnetic field, the computed response is remarkably stable to errors in the measured Green's functions. This motivates investigation of the noise robustness of our approach and its suitability for implementation on noisy intermediate scale quantum (NISQ) quantum computers and simulators. For the examples, presented in this manuscript, the degree of freedom encoding the electromagnetic field did not have be bosonic since we only considered one photon states; such simulations could be performed solely by encoding in internal state of ions. However, for multi-photon interactions, the bosonic/harmonic nature of the vibrational modes becomes essential. In this work we have focused on the trapped-ion platform for concreteness. However, any quantum simulation platform with controllable bosonic modes with tunable coupling to other localized (e.g., qubit) degrees of freedom could implement our scheme.

\begin{acknowledgments}
This work was supported by the U.S. Department of Energy, Office of Science,  Office of Basic Energy Sciences under the Materials and Chemical Sciences Research for Quantum Information Research program. The final stages of this work were also supported by the U.S. Department of Energy, Office of Science, Office of Advanced Scientific Computing Research, under the Quantum Computing Application Teams (QCAT) program.

This article has been authored by an employee of National Technology \& Engineering Solutions of Sandia, LLC under Contract No. DE-NA0003525 with the U.S. Department of Energy (DOE). The employee owns all right, title and interest in and to the article and is solely responsible for its contents. The United States Government retains and the publisher, by accepting the article for publication, acknowledges that the United States Government retains a non-exclusive, paid-up, irrevocable, world-wide license to publish or reproduce the published form of this article or allow others to do so, for United States Government purposes. The DOE will provide public access to these results of federally sponsored research in accordance with the DOE Public Access Plan \url{https://www.energy.gov/downloads/doe-public-access-plan.}
\end{acknowledgments}

\bibliographystyle{unsrt}
\bibliography{light_absorption}

\appendix
\section{Derivation of ancilla-assisted dissipation}
\label{app:diss}
In order to engineer the correlated dissipation induced by coupling of the degrees of freedom in the matter subsystem to a broadband EM field, we introduce an auxiliary ion and isolate two internal states within it governed by a two-level system with Hamiltonian $\hat{H}_\ax=\frac{\omega_\ax}{2}(\hat{1}-\hat{\sigma}_\ax^z)$. Then we assume that the excited state is subject to a fast decay process: $\mathcal{D}[\hat{X}]$, with $\hat{X}=\chi \hat{\sigma}_\ax^-$. This could be engineered through optical pumping to a higher lying state, for example \cite{Barreiro_Blatt_2011}, which is a generally useful entropy reduction resource in quantum simulation \cite{herdman_2010, metcalf_2020}. Then, consider the combined system governed by the model:
\begin{flalign}
	\bar{G}_M(t)&=\expn{-i\scomm{\hat{H}_M}t}\nnl
	\bar{G}_\ax(t)&=\expn{(-i\scomm{\hat{H}_\ax}-\frac{1}{2}\sacomm{\hat{X}\dg\hat{X}})t}\nnl
	\bar{G}_0(t)&=\bar{G}_M(t)\otimes\bar{G}_\ax(t)\nnl
	\hat{H}_{M-\ax}&=\hat{J}\hat{\sigma}^+_\ax+h.c.=\sum_iJ_i\hat{\sigma}^-_i\hat{\sigma}^+_\ax+h.c.\nnl
	\bar{F}_{M-\ax}&=-i\scomm{\hat{H}_{M-\ax}}+\sop{\hat{X}}{\hat{X}}
\end{flalign}
Here, the term $\hat{H}_{M,\ax}$ describes the coherent coupling between the optically active states in the matter subsystem and the auxiliary ion's internal states, which are engineered using the M{\o}lmer-S{\o}renson interaction, as in Eq. (\ref{eq:gen_ion_ham}). Since we want to understand the effects of coupling to the auxilliary ion we do not include the mode coupling and any additional decoherence terms on the matter subsystem in the above for simplicity. The intuition here is that this coherent coupling to a fast decaying level will allow the matter states to ``inherit'' some amount of dissipative decay dynamics, and this amount can be tuned by choice of the coupling parameters $J_i$ for each excited state in the matter subsystem indexed by $i$. The coupling terms are assumed to be small, and thus in the following, we will expand perturbatively in their magnitude.
\begin{flalign*}
	\bar{G}(t)&=\bar{G}_0(t)+\int^t_0 dt'\bar{G}_0(t-t')\bar{F}_{M-\ax}\bar{G}_0(t')\\
	&+\int^t_0 dt'\bar{G}_0(t-t')\bar{F}_{M-\ax}\int^{t'}_0 dt''\bar{G}_0(t'-t'')\bar{F}_{M-\ax}\bar{G}_0(t'')+...\\
\end{flalign*}
If we are only interested in the impact on the matter subsystem states, we keep only relevant terms that describe transfer both to and from the auxiliary state in this expansion.  For an initial state where the auxilliary system is in its ground state,
\begin{flalign*}
	\sbra{0_\ax}{0_\ax}&\bar{G}(t)\sket{0_\ax}{0_\ax}\\
	&=\bar{G}_M(t)-\int^t_0 dt'\int^{t'}_0 dt''\bar{\mathcal{L}}_+(t,t')\bar{\mathcal{L}}_-(t',t'')\bar{G}_M(t'')\\
	&+\int^t_0 dt'\int^{t'}_0 dt''\int^{t''}_0 dt'''\sop{\hat{X}}{\hat{X}}\bar{\mathcal{L}}\dg_-(t',t'')\bar{\mathcal{L}}_-(t'',t''')\bar{G}_M(t''')\\
	&+\int^t_0 dt'\int^{t'}_0 dt''\int^{t''}_0dt'''\int^{t'''}_0 dt'''' \bar{\mathcal{L}}\dg_+(t,t')\bar{\mathcal{L}}_+(t',t'')\bar{\mathcal{L}}\dg_-(t'',t''')\bar{\mathcal{L}}_-(t''',t'''')\bar{G}_M(t'''') +h.c.+\dots\\
	&=\bar{G}_M(t)-\int^t_0 dt'\int^{t'}_0 dt''\bar{G}_M(t-t')\sop{\hat{J}\dg}{\iop}\\
	&\times\bar{G}_M(t'-t'')\sop{\hat{J}}{\iop}\expn{-\frac{\chi^2}{2}(t'-t'')}\bar{G}_M(t'')+\int^t_0 dt'\int^{t'}_0 dt''\int^{t''}_0 dt'''\chi^2\expn{-\chi^2(t'-t'')}\sop{\iop}{\hat{J}}\\
	&\bar{G}_M(t''-t''')\expn{-\frac{\chi^2}{2}(t''-t''')}\sop{\hat{J}}{\iop}\bar{G}_M(t''')\\
	&+\int^t_0 dt'\int^{t'}_0 dt''\int^{t''}_0dt'''\int^{t'''}_0 dt'''' \bar{G}_M(t-t')\sop{\iop}{\hat{J}\dg}\bar{G}_M(t'-t'')\expn{-\frac{\chi^2}{2}(t'-t'')}\sop{\hat{J}\dg}{\iop}\\
	&\times\expn{-\chi^2(t''-t''')}\sop{\iop}{\hat{J}}\bar{G}_M(t'''-t'''')\expn{-\frac{\chi^2}{2}(t'''-t'''')}\sop{\hat{J}}{\iop}\bar{G}_M(t'''') +h.c.+\dots\\
\end{flalign*}
with $\bar{\mathcal{L}}_+(t,t')=\bar{G}_0(t-t')\sop{\hat{J}^\dg\hat{\sigma}^-_\ax}{\iop}$ and $\bar{\mathcal{L}}_-(t,t')=\bar{G}_0(t-t')\sop{\hat{J}\hat{\sigma}^+_\ax}{\iop}$
If we assume $\chi$ is large compared to the other processes in the system, we find that there are three effective first order terms
\begin{flalign}
	\sbra{0_\ax}{0_\ax}&\bar{G}(t)\sket{0_\ax}{0_\ax}\nnl
	&\approx\bar{G}_M(t)-\frac{2}{\chi^2}\int^t_0 dt'\bar{G}_M(t-t')\sop{\hat{J}\dg\hat{J}}{\iop}\bar{G}_M(t')\nnl
	&+\frac{2}{\chi^2}\int^t_0 dt'\sop{\hat{J}}{\hat{J}}\bar{G}_M(t')\nnl
	&+\frac{4}{\chi^6}\int^t_0 dt'\bar{G}_M(t-t')\sop{\hat{J}\dg\hat{J}}{\hat{J}\dg\hat{J}}\bar{G}_M(t') +h.c.+\dots\label{eq:aux_expans}
\end{flalign}
The contribution of the last term vanishes in the large $\chi$ limit.  The above then becomes
\begin{flalign*}
	\sbra{0_\ax}{0_\ax}&\bar{G}(t)\sket{0_\ax}{0_\ax}\approx\expn{\left(-i\bar{\mathcal{H}}_M+\frac{4}{\chi^2}\left[\sop{\hat{J}}{\hat{J}}-\frac{1}{2}\sacomm{\hat{J}\dg\hat{J}}\right]\right)t}
\end{flalign*}
We see that the impact of the auxiliary state dynamics on the system is equivalent to a decay process for the system that would enter the master equation as $\mathcal{D}[\frac{2\hat{J}}{\chi}]$. Therefore, we can engineer the non-local decay terms necessary to model the dissipative effect of coupling to a continuum of modes by tuning the coherent couplings in $\hat{J}$ such that $\frac{2\hat{J}}{\chi} = \hat{L}$.

\end{document}